\documentclass[ preprint,
 amsmath,amssymb,
 aip,
]{revtex4-2}

\usepackage{graphicx}\usepackage{dcolumn}\usepackage{bm}
\usepackage[T1]{fontenc}    \usepackage{hyperref}       \usepackage{url}            \usepackage{booktabs}       \usepackage{amsfonts}       \usepackage{nicefrac}       \usepackage{microtype}      \usepackage{xcolor}         \usepackage{graphicx}
\usepackage{amsmath}

\DeclareMathOperator*{\argmin}{arg\,min}

\begin{document}
\title{Learning atomic forces from uncertainty-calibrated adversarial attacks
}
\author{Henrique Musseli Cezar}
\affiliation{Hylleraas Centre for Quantum Molecular Sciences and Department of Chemistry, University of Oslo, PO Box 1033 Blindern, 0315 Oslo, Norway}

\author{Tilmann Bodenstein}
\affiliation{Hylleraas Centre for Quantum Molecular Sciences and Department of Chemistry, University of Oslo, PO Box 1033 Blindern, 0315 Oslo, Norway}

\author{Henrik Andersen Sveinsson}
\affiliation{The Njord Centre, Department of Physics, University of Oslo, PO Box 1048 Blindern, 0316 Oslo, Norway}

\author{Morten Ledum}
\affiliation{Hylleraas Centre for Quantum Molecular Sciences and Department of Chemistry, University of Oslo, PO Box 1033 Blindern, 0315 Oslo, Norway}

\author{Simen Reine}
\affiliation{Hylleraas Centre for Quantum Molecular Sciences and Department of Chemistry, University of Oslo, PO Box 1033 Blindern, 0315 Oslo, Norway}

\author{Sigbjørn Løland Bore}
\email{s.l.bore@kjemi.uio.no}
\affiliation{Hylleraas Centre for Quantum Molecular Sciences and Department of Chemistry, University of Oslo, PO Box 1033 Blindern, 0315 Oslo, Norway}

\begin{abstract}
Adversarial approaches, which intentionally challenge machine learning models by generating difficult examples, are increasingly being adopted to improve machine learning interatomic potentials (MLIPs).
While already providing great practical value, little is known about the actual prediction errors of MLIPs on adversarial structures and whether these errors can be controlled. 
We propose the Calibrated Adversarial Geometry Optimization (CAGO) algorithm to discover adversarial structures with user-assigned errors. Through uncertainty calibration, the estimated uncertainty of MLIPs is unified with real errors. 
By performing geometry optimization for calibrated uncertainty, we reach adversarial structures with the user-assigned target MLIP prediction error.
Integrating with active learning pipelines, we benchmark CAGO, demonstrating stable MLIPs that systematically converge structural, dynamical, and thermodynamical properties for liquid water and water adsorption in a metal-organic framework within only hundreds of training structures, where previously many thousands were typically required. 
\end{abstract}
\maketitle
\clearpage
\section{Introduction}
By representing the potential energy surface of atoms with neural networks, machine learning interatomic potentials (MLIPs) can be trained to predict the outcomes of costly quantum mechanical calculations in milliseconds instead of hours. With pioneering MLIPs, such as Behler-Parinello neural networks, structural and thermodynamic properties are already well captured.\cite{behler_generalized_2007} The latest equivariant message passing neural network approaches, such as NequIP,\cite{batzner_e3-equivariant_2022} Allegro,\cite{musaelian_learning_2023} and MACE\cite{NEURIPS2022_4a36c3c5} have further significantly reduced prediction errors by nearly an order of magnitude. These approaches typically achieve training and test set errors far lower than typical errors associated with the details of the underlying quantum mechanical calculations they are trained on, for example, basis set truncation and functional choice in the case of Density Functional Theory (DFT). Although MLIPs rely on the inductive bias of short-range interactions, their accuracy makes these methods, in principle, capable of accurately representing the energy landscape of chemically complex systems.

Parameterization of reliable MLIPs remains a challenge, often requiring significant human time and trial-and-error experimentation. The challenge in developing a reliable MLIP presents a chicken-and-egg scenario. On the one hand, the ultimate goal of MLIPs is to explore the unknown, reaching beyond the capabilities of ab initio MD, extending both to longer timescales and to larger length scales. On the other hand, the phase space covered by the training set is limited by the computational cost of ab initio calculations. As a result, all applications rely to varying degrees on the MLIP's ability to generalize to structures outside the training set. This ability to generalize is not just a property of the MLIP but an interaction between the MLIP and the training set. This dependency can become very problematic, as highlighted in recent studies\cite{zhai_short_2023,fu2023forces} which demonstrate that MLIPs, in some situations, struggle to provide stable dynamics, accurate sampling, or reproduce the underlying physics.  

Since current MLIP approaches achieve very good training set and validation set accuracy, this problem ultimately arises due to shifting features (molecular structures in our case) from training to production. When the MLIP is used in practice, the molecular structures are different from the structures in the training data. This phenomenon is known in a broader statistics context as a covariate shift.\cite{shimodaira_improving_2000} Active learning approaches have become the go-to solution for reducing covariate shifts in structures that typically occur during molecular dynamics (MD) sampling with MLIP-based models. The central idea is to use the MLIP to extend the training set by sampling structures that the MLIP encounters during production.\cite{yang2024machine} In practice, this is typically achieved by performing iterations of MD sampling with the MLIP, adding new structures until the MLIP model reaches specific convergence criteria, such as simulation stability or the accurate reproduction of structural and thermodynamic properties of reference ab initio simulations. 

To avoid unnecessary costly reference calculations on structures already well represented in the training set, active learning procedures typically employ uncertainty quantification to select structures that enhance the training data, i.e., those with high prediction uncertainty. There are various approaches to uncertainty estimation, including Gaussian Process regression,\cite{vandermause_--fly_2020,xie_uncertainty-aware_2023} dropout in neural networks,\cite{wen_uncertainty_2020} and committees of MLIPs.\cite{NIPS1994_b8c37e33} While the first two methods rely on specific MLIP architectures, the committee approach is broadly applicable, only requiring computing the variance of a property between multiple models trained on different training sets or seeds, for example, through bootstrapping. However, it is not without drawbacks: The members of the MLIP committee can potentially all agree on an incorrect answer. Furthermore, training and running multiple models adds to the computational cost. Despite these issues, the committee approach remains widely used due to its simplicity and ease of implementation, with many well-established active learning softwares using it.\cite{smith_less_2018,zhang_dp-gen_2020,schran_machine_2021,vandenhaute_machine_2023}

For uncertainty estimation to effectively extend the selection of new training set structures, it is essential to offer a diverse pool of candidate structures that expand the phase space of the existing training set. The typical strategy involves MD sampling with MLIPs across various thermodynamic conditions. The inclusion of structures that represent rare events through enhanced sampling techniques is also increasingly being recognized as important.\cite{yang_using_2022,vandenhaute_machine_2023} Although MLIP-based sampling with MD is a simple approach suitable for automated active learning pipelines to develop models, it has some weaknesses. These include prolonged correlation times, leading to high inter-structure correlation, and the inherent Boltzmann bias toward low-energy structures. Therefore, when the standard MLIP sampling approach fails, there are numerous pragmatic solutions to target more diverse structures, such as sampling at high temperatures and pressures, normal-mode sampling, or perturbing structures via random atomic displacements.\cite{smith2018less,zhang_dp-gen_2020} While these methods offer structures contrasting those of standard MD simulations, they often result in high-energy structures with atomic overlap. The result is, therefore, an undesirable compromise between including structures with high prediction uncertainty and structures with significant forces from atomic overlap. The latter can reduce the overall accuracy of the MLIP.\cite{zeng2020complex} In addition, such approaches tend to localize the prediction uncertainty to a few atoms, which is problematic for large molecular assemblies. It stands to reason that a few carefully optimized structures can offer the same learning content as many highly correlated structures.

In machine learning for image classification problems, adversarial approaches provide a systematic framework for building more robust models by training against examples that aim to trick the model.\cite{goodfellow2016deep} This approach has also been applied to the active learning of MLIPs. To the best of our knowledge, this idea was first introduced by Cubuk et al.\cite{cubuk_adversarial_nodate}\ to move atoms toward high prediction uncertainty for the potential energy. A similar idea was also applied in ref.~\citenum{kulichenko_uncertainty-driven_2023} using a bias towards high uncertainty with metadynamics simulations. Such approaches are algorithmically elegant by requiring only a standard MLIP single point force calculation. However, energy uncertainty leaves 3$N$+6 labels of learning content associated with gradients and virials up to chance. In this regard, the Bombarelli Group extended the approach of Cubuk to include prediction uncertainty of forces.\cite{schwalbe-koda_differentiable_2021} Using a differentiable MLIP architecture, they performed an adversarial active learning approach capable of discovering structures with high force uncertainty, reaching robust models for challenging systems including zeolite and alanine molecules. This work was also recently extended to non-periodic system reference calculations,\cite{roy2024learning} and force uncertainty-driven dynamics.\cite{zaverkin2024uncertainty} 

\begin{figure*}
    \includegraphics[width=0.9\textwidth]{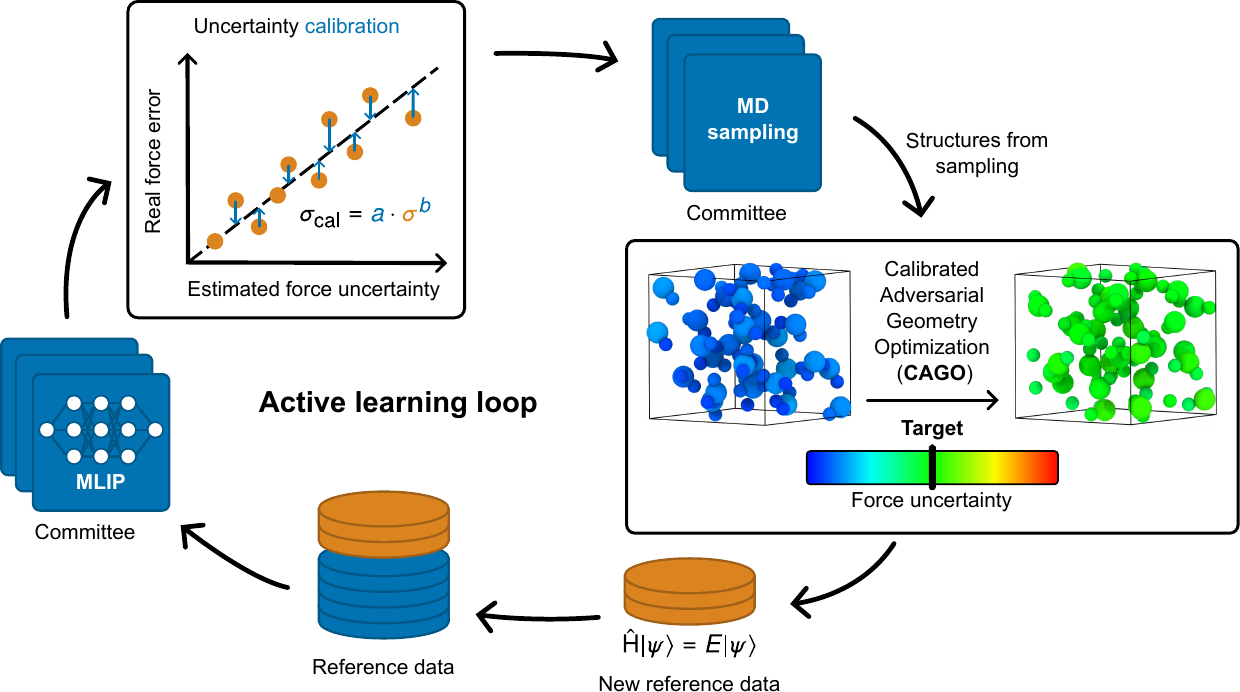}
    \caption{
    Active learning loop with calibrated adversarial geometry optimization (CAGO). Uncertainty calibration, the determination of calibration parameters, is performed using training data and a committee of MLIPs. The CAGO algorithm is applied to the MD-sampled structures from the current iteration of the MLIPs to obtain new structures. These structures are subsequently used in reference ab initio calculations and added to the training set.
    }
    \label{fig:al}
\end{figure*}

While such adversarial approaches have demonstrated great potential in active learning, very little attention has so far been put towards quantifying the errors of MLIPs on the adversarial structures. As such, a fundamental question remains: To what extent can we control the actual errors of the MLIP on the adversarial structures? To answer this question, we have developed the Calibrated Adversarial Geometry Optimization (CAGO) algorithm, which aims to discover new adversarial structures with target force errors preassigned by the user (Figure~\ref{fig:al}). To unify estimated prediction uncertainties with real errors, we perform uncertainty calibration. To control the MLIP prediction errors on adversarial structures, we optimize structures to reach moderate target errors. These structures are within the range of validity of the uncertainty calibration while being challenging structures for the MLIPs, from which they can learn. To demonstrate the usefulness of our approach, we integrate CAGO into an active learning framework, as shown in Figure~\ref{fig:al}, enabling us to learn liquid-water dynamics and water adsorption in metal-organic frameworks from small datasets.

\section{Results}
\subsection{Uncertainty quantification and calibration}
MLIPs predict quantities $y$, such as forces, energies, and virials, depending on the structure $\mathbf x$, and we would like to estimate the root mean square error $\sigma_{\text{rmse}}$ of the predictions $\hat y$ with respect to a ground truth reference $y_{\text{ref}}$:
\begin{equation}
    \sigma_{\text{rmse}}^2(\mathbf x)=\frac 1M \sum^M_{m=1}\left|\hat{y}^m-y_{\text{ref}}\right|^2,
\end{equation}
where $m$ denotes one of the $M$ models. In the committee approach, this error is estimated from the standard deviation of predictions:
\begin{equation}
    \hat\sigma^2 = \frac{1}{M-1}\sum_{m=1}^M\left(\hat{y}^m-\bar y\right)^2,
\end{equation}
where $\bar y$ denotes the committee mean. In practice, the committee uncertainty estimates $\hat\sigma$ typically underestimates the actual prediction error  $\sigma_{\text{rmse}}$. To achieve a statistically correct uncertainty estimate, we employ uncertainty calibration, following the procedure introduced in ref.~\citenum{palmer_calibration_2022}, where the uncertainty estimate $\sigma$ is considered well-calibrated when the ratio
\begin{equation}
    r = \frac{\hat y-y_{\text{ref}}}{\sigma}
\end{equation}
is normally distributed with bias zero, and the standard deviation is equal to 1. Many different uncertainty calibration schemes have been proposed.\cite{musil_fast_2019,imbalzano_uncertainty_2021} Here, we opt for a power law calibration strategy:\cite{musil_fast_2019}
\begin{equation}\label{eq:error-cal}
    \sigma_{\text{cal}}=a\cdot\hat \sigma^b,
\end{equation}
where $a$ and $b$ are determined by optimizing the negative log-likelihood that $\hat y-y_{\text{ref}}$ was drawn from a normal distribution with zero bias and standard deviation in eq.~\eqref{eq:error-cal} over structures $\mathbf x:$\cite{palmer_calibration_2022}
\begin{equation}\label{eq:log-likelihood}    a,\,b=\argmin_{a',\,b'}\sum_{\mathbf{x}} \left[2\pi+\ln\left(a'\hat \sigma^{b'}\right)^2+\frac{\left|\hat y(\mathbf x)-y_{\text{ref}(\mathbf x)}\right|^2}{\left(a'\hat \sigma^{b'}\right)^2}\right].
\end{equation}

\subsection{Adversarial structures}
Like in ref.~\citenum{schwalbe-koda_differentiable_2021}, we create adversarial attacks by optimizing the structure $\mathbf x$ according to a fitness function $\mathcal L$. However, instead of maximizing the committee uncertainty estimate $\hat\sigma$, we optimize calibrated prediction uncertainties $\sigma_\text{cal}$ towards the error target $\delta$:
\begin{equation}
    \mathcal L(\sigma_{\text{cal}})=\left(\sigma_{\text{cal}}(\mathbf x)-\delta\right)^2.
\end{equation}
By targeting a specific prediction uncertainty $\delta$, we aim to push the predictions $\hat y$ outside the MLIPs comfort zone, where the error is considerably higher than the training set error. This ensures that the resulting structures contain information that expands the training set while at the same time maintaining the consistency between real errors and estimated errors.

\subsection{Biasing adversarial structures}
In addition to the prediction uncertainty of adversarial structures, it can be desirable to bias the adversarial structural properties toward a target value for $y_{\text{bias}}$, e.g., certain pressures or force magnitudes less than a specified threshold. This can be achieved by supplementing the fitness function by:
\begin{equation}
    \mathcal L_{\text{bias}}(\bar y)=l_{\text{bias}}\left(\bar y(\mathbf x)-y_{\text{bias}}\right)^2,
\end{equation}
where $l_{\text{bias}}$ is a prefactor determining how strictly the bias is applied.  

\subsection{Force-based calibrated adversarial geometry optimization}
Throughout the rest of this paper, we will consider force-based adversarial geometry optimization, with the molecular structure $\mathbf x$ being determined by the optimization problem:
\begin{equation}
    \mathbf{x} = \argmin_\mathbf{x}\sum\limits_{i=1}^{N_\text{atoms}} \left((\hat\sigma_{\mathbf F_i}(\mathbf x)-\delta)^2+l_{\text{bias}}\left|\bar F_i(\mathbf x)\right|^2\right),
\label{eq:opt-problem}
\end{equation}
where $N_\text{atoms}$ is the number of atoms, $\hat\sigma_{\mathbf F_i}$ is given by
\begin{equation}
\hat\sigma_{\mathbf F_i}=\sqrt{\hat\sigma_{|F_{ix}|}^2+\hat\sigma_{|F_{iy}|}^2+\hat\sigma^2_{|F_{iz}|}},
\end{equation}
and $|\bar F_i(\mathbf x)|$ denotes the norm of the average force on atom $i$.

\subsection{Calibrated adversarial geometry optimization to target error}\label{sec:res-cali}
\begin{figure}[!htb]
    \centering
    \includegraphics[width=0.49\textwidth]{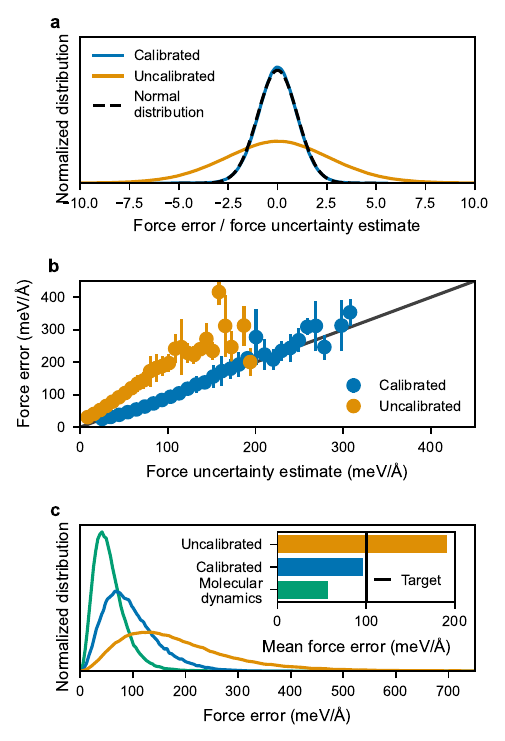}
    \caption{Uncertainty calibration and error statistics on adversarial structures. {\bf a.} Comparison of error statistics of forces predicted by a committee of models divided by estimated prediction uncertainty against normal distribution. {\bf b.} Force errors for a committee of models vs.\,estimated uncertainty. {\bf c.} Force error statistics for structures from MD, CAGO with calibrated uncertainty, and CAGO without uncertainty-calibration, with associated estimated mean errors in the inset figure.}
    \label{fig:adv-illu}
\end{figure}

We start by considering the error of force predictions on liquid water from a committee of 20 MLIPs trained on a synthetic training set of DNN@MB-pol\cite{bore_realistic_2023} MD trajectories (see Methods~\ref{sec:cago-benchmark} for more details). Figure~\ref{fig:adv-illu}a compares the ratio of the distribution error and the estimated calibrated/uncalibrated uncertainties against the normal distribution from ideal error calibration.\cite{palmer_calibration_2022} While both curves show a bell curve, the wider shape of the force error\textendash uncalibrated uncertainty ratio shows that the uncalibrated uncertainty significantly underestimates the true uncertainty. In contrast, the calibrated uncertainty leads to an almost perfect agreement. In the Supplementary Information, we show that this is also the case for energy and virial predictions. The near-perfect error statistics for the force errors of the calibrated uncertainty is reflected in the one-to-one correlation between the mean force error and the force uncertainty estimate in Figure~\ref{fig:adv-illu}b. For high force errors, the correlations are good, albeit more noisy than in the low-force-errors regime. High-error samples are inherently less frequent than low-error samples. This means the uncertainty calibration has access to less data for high-error samples and is therefore expected to be less accurate in this case than in the low-error regime. As such, the force error versus the force uncertainty can act as a heuristic to determine the validity range for calibrated uncertainty estimation, in this case, up to about $\sim200$~meV/\AA.

Next, we perform CAGO to solve equation~\eqref{eq:opt-problem} without a bias-term by geometry optimization using numerical finite-difference gradients for 40 structures (see Figure~\ref{fig:al}, all structures reaching the target force error indicated in green). Figure~\ref{fig:adv-illu}c reports the corresponding error statistics, i.e., individual model forces versus reference DNN@MB-pol forces. The error statistics for adversarial structures for CAGO with calibrated uncertainties lead to a mean error close to the target error, whereas the uncalibrated case has errors about twice as high as the target. The potential usefulness of such calibrated adversarial structures can be understood when compared to ordinary MD structures, which have about half the error and would, therefore, not provide the same level of learning content if added to the training set.
\begin{figure}[!htb]
    \centering
    \includegraphics[width=1\textwidth]{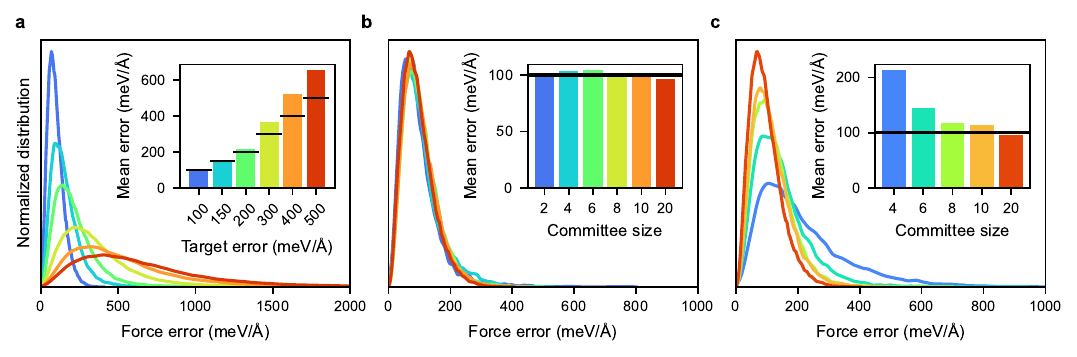}
    \caption{Benchmark on the effect of CAGO algorithm hyperparameters. The panels show the distribution of MLIP force errors on adversarial structures, with corresponding mean errors in the inset of each panel. Line legends and colors are specified by ticks in inset figures. \textbf{a.} Force error statistics for different error targets. \textbf{b.} Force error statistics for adversarial structures optimized for a target 100 meV/\AA\ uncertainty with different committee sizes, using uncertainty calibration from a committee of 20 MLPs. \textbf{c.} Force error statistics for adversarial structures optimized for a target 100~meV/\AA\ uncertainty, considering uncertainty calibration performed with its respective committee size.}
    \label{fig:benchmark}
\end{figure}

The CAGO algorithm has several hyperparameters that can be adjusted in accordance with user goals and are benchmarked in Figure~\ref{fig:benchmark}. Figure~\ref{fig:benchmark}a reports the error statistics for different error targets $\delta$ in equation~\eqref{eq:opt-problem}. By tuning this parameter, the real error of adversarial structures can be controlled. A clear trend here is that CAGO is more accurate when targets have moderate values, in this case, up to about 200~meV/\AA, while for higher values, the real errors are higher than the target errors. This is in line with our observations from Figure~\ref{fig:adv-illu}b, where above 200~meV/\AA\ the uncertainty calibration is less accurate.

When performing CAGO, using a small committee of models is desirable due to the linear increase in computational cost with committee size. Interestingly, Figure~\ref{fig:benchmark}b shows that the error statistics are not very sensitive to committee size, with a few models sufficing to reach close to the target error. On the contrary, Figure~\ref{fig:benchmark}c shows how committee size is critical for achieving reliable calibration, where adversarial structures from small committees calibrated with the same number of models do not hit their target. In this case, a committee of around 10 models is necessary to reach the target error, but this may be system- and MLIP-architecture-dependent. This indicates that using many models to calibrate the uncertainties (Figures~\ref{fig:benchmark}c) and performing CAGO using a few models (Figures~\ref{fig:benchmark}b) is a good heuristics for performing CAGO reliably and efficiently. CAGO can also be performed with bias terms while maintaining the target error, as demonstrated by our benchmark reported in the Supplementary Information.

\subsection{Learning liquid water from a single structure}
\begin{figure*}[!htb]
    \centering
    \includegraphics[width=0.99\linewidth]{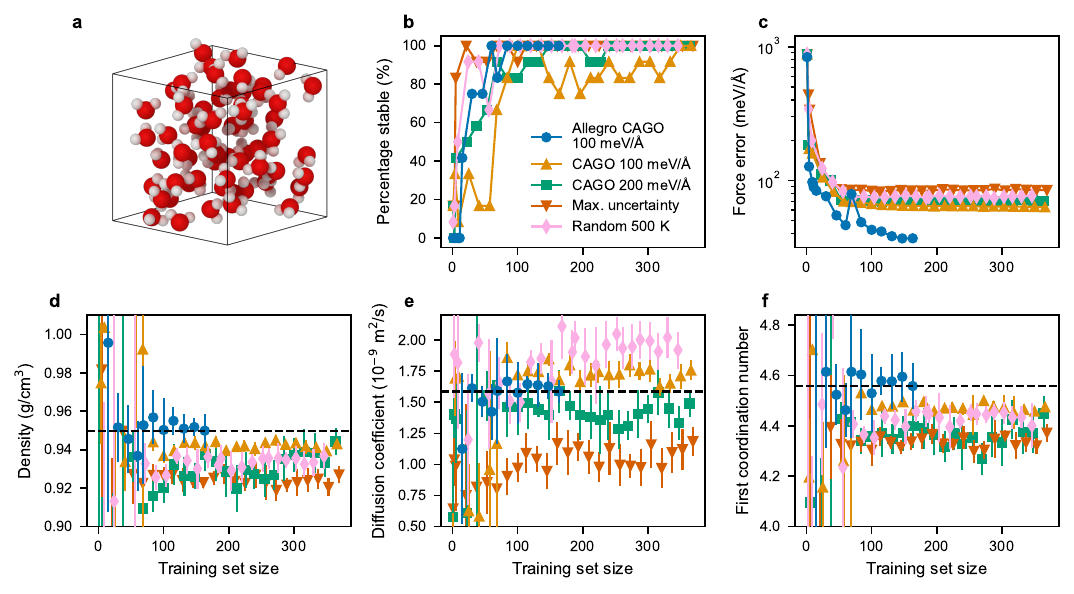}
    \caption{Benchmark of convergence of liquid-water properties for CAGO-based and standard active learning starting out from a single structure. We report data for the Allegro MLIP with target error 100~meV/\AA\ and for DeePMD with target error 100~meV/\AA, 200~meV/\AA, as well as picking maximum uncertainty structure from MD and random structure sampling from 500~K MD simulations (see Methods~\ref{sec:stability} for more details). Each point in every graph has been computed from 12 models using the subset of stable MLIPs, with three replica MD simulations lasting 2~ns each, for a total of 540 Allegro and 3744 DeepMD simulations. The error bars are the standard deviation between the different committee members, and the horizontal lines correspond to the average final value of the property for Allegro CAGO. \textbf{a.} The liquid water box structure that was used to start the active learning training. Oxygen is represented in red, with hydrogen in white. \textbf{b.} Percentage of stable MLIPs in accordance with the stability criteria in Methods~\ref{sec:stability}. \textbf{c.} Average force error of MLIPs on the liquid water structures. \textbf{d.} Mean liquid water density. \textbf{e.} Self-diffusion coefficient. \textbf{f.} First coordination number (integral of the first radial distribution function peak).}
    \label{fig:water-deepmd-allegro}
\end{figure*}

An MLIP is only as good as its reference data. Therefore, creating a training set with as high-quality references as possible is desirable. However, higher quality is generally associated with higher computational cost. In particular, density-corrected DFT (DC-DFT) and density-corrected R$^2$SCAN (DC-R$^2$SCAN) achieve excellent correspondence in energies with respect to coupled cluster methods.\cite{dasgupta_elevating_2021,dasgupta_nuclear_2024} However, DC-DFT and DC-R${}^2$SCAN are rather expensive compared to ordinary functionals, making the standard route of starting from a training set of ab initio MD trajectories prohibitively expensive. It is therefore important to have a method that can start the active learning process with as few training structures as possible. Taking this to the extreme, here we use active learning to learn an MLIP for DC-R$^2$SCAN starting from only a single structure of liquid water (Figure~\ref{fig:water-deepmd-allegro}a). Figure~\ref{fig:water-deepmd-allegro} reports our benchmark on the performance of MLIP iterations during CAGO-based active learning for Allegro with error target 100~meV/\AA\ (Allegro CAGO 100~meV/\AA), and, for DeePMD, with targets 100~meV/\AA\ and 200~meV/\AA\ (CAGO 100~meV/\AA\ and 200~meV/\AA), standard active learning selecting maximum uncertainty structures from MD (max.\ uncertainty), and sampling random structures from MD at 500~K (random 500~K), please refer to Methods~\ref{sec:stability} for additional details. Note that for the first two iterations of CAGO-based active learning, we do CAGO from the first starting geometry, and not from MD sampled geometries (see Methods~\ref{sec:stability} for additional details).

Figure~\ref{fig:water-deepmd-allegro}b reports the percentage of stable models. At around 80 training-set structures, all Allegro models become stable, while the CAGO-based DeePMD models achieve stability at around 220 or 350 for the 200~meV/\AA\ and 100~meV/\AA\ targets, respectively. The slower convergence of DeePMD is in line with past research, which shows that MLIPs based on equivariant layers are not only more accurate but also more data-efficient,\cite{batzner_e3-equivariant_2022,musaelian_learning_2023} with similar equivariant NequIP also being benchmarked to be more stable than DeePMD.\cite{fu2023forces} However, DeePMD, with our CAGO structures, performs rather well compared to its data-hungry reputation.\cite{batzner_e3-equivariant_2022} To put this into context, ref.~\citenum{fu2023forces} used 10\,000 structures and achieved stability for an average of 0.247 nanoseconds under $NVT$ conditions with a simple water force-field. In contrast, we achieve 100~\% stability for two nanoseconds under the more challenging $NPT$ conditions for a complex many-body ab initio method. Interestingly, active learning using more standard approaches (max.\ uncertainty and random 500~K) achieves stability quicker than CAGO-based active learning. This could be due to the sampling yielding higher uncertainty structures than CAGO, which targets moderate force errors. This would be in line with the previously observed trend of higher target errors leading to more stable models. 

To benchmark how well CAGO-based active learning performs for energetics, we report in Figure~\ref{fig:water-deepmd-allegro}c the average force error for 100 undistorted liquid water MD structures. In agreement with the literature,\cite{musaelian_learning_2023} the Allegro force errors are considerably smaller than those of DeePMD, with Allegro CAGO 100 meV/\AA\ converging to 37~meV/\AA, and DeePMD CAGO 100 meV/\AA\ and 200 meV/\AA\ converging to 63~meV/\AA\ and 70~meV/\AA, respectively. Similar force error magnitudes were also reported in ref.~\citenum{batzner_e3-equivariant_2022}. It is noteworthy that while CAGO distorts the structures from MD during training, CAGO improves force errors of MLIPs on undistorted liquid MD structures. Although stability for DeePMD is achieved with fewer training structures for standard active learning approaches, the force errors are higher: max.\ uncertainty gives an error of 85~meV/\AA, (approximately 35~\% higher than CAGO 100~meV/\AA), and random 500~K gives an error of  77~meV/\AA\ (approximately 22~\% higher than CAGO 100~meV/\AA). 

Next, in Figures~\ref{fig:water-deepmd-allegro}d, e, and f, we benchmark the thermodynamical, dynamical, and structural properties of liquid water, respectively. As with the stability, for all properties, Allegro converges around 80 structures, with DeePMD following closely. Overall, DeePMD exhibits larger variability for these properties. This is consistent with past benchmark studies on MB-pol-based DeePMD, which, even for huge training sets, struggle to achieve the correct density,\cite{zhai_short_2023,zhai2024many} whereas Allegro gets it spot on.\cite{maxson2024transferable} Nonetheless, we only observe variability within a few percent, and CAGO-based DeepMD models are within one standard deviation of the final Allegro iteration. For DeePMD, comparing against more standard active learning approaches (max.\ uncertainty and random 500~K), CAGO with 100 meV/\AA\ performs best, followed by CAGO with 200 meV/\AA, followed closely by random 500~K, and max.\ uncertainty. This order of accuracy is the same as the order of accuracy for the energetics as reported in Figure~\ref{fig:water-deepmd-allegro}c (see Table~S3 in the Supplementary Information for the numbers of the final models).  

\subsection{Learning water-adsorption of a metal-organic framework}
\begin{figure*}[!htb]
    \centering
\includegraphics[width=1.\textwidth]{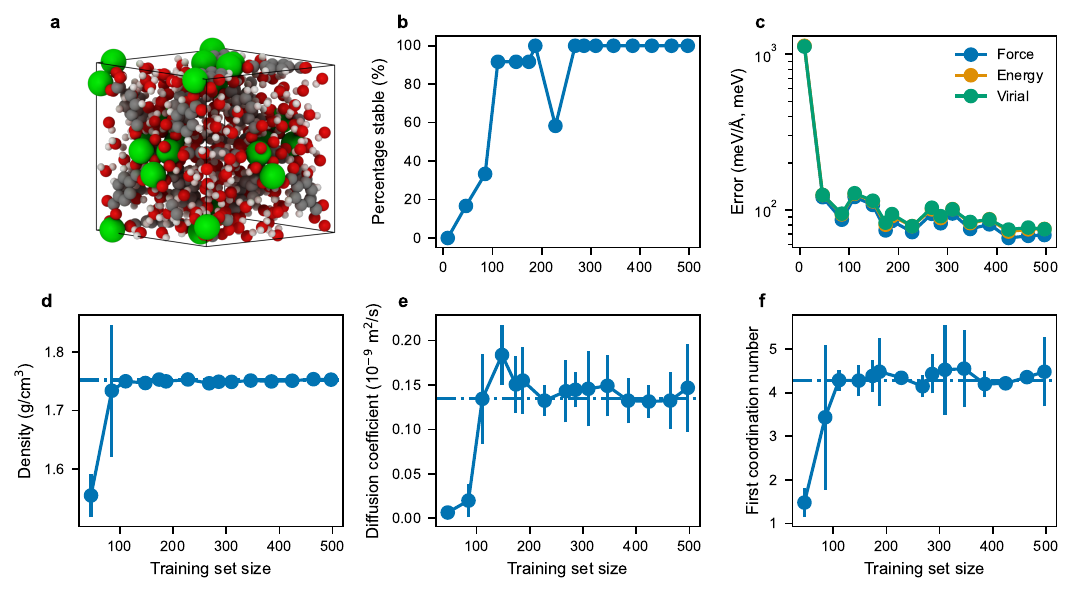}
    \caption{Benchmark of convergence of properties for water-adsorption in MOF for CAGO-based active learning for DC-R$^2$SCAN, with Allegro MLIPs. When presented, the error bars are the standard deviation of the measurements of each committee member at that iteration. Each point in the graphs has been computed from 12 models using the subset of stable MLIPs, with three replica simulations each, for a total of 576 two nanoseconds MD simulations. The error bars and horizontal lines hold the same meaning as in Figure~\ref{fig:water-deepmd-allegro}.
\textbf{a.} Water MOF system used in this benchmark. The color coding for the atoms is red, white, gray, and green for oxygen, hydrogen, carbon, and zirconium, respectively. \textbf{b.} Percentage MLIPs fulfilling stability criteria in Methods~\ref{sec:stability}. \textbf{c.} Average force, energy, and virials error on MOF/water structures. \textbf{d.} Mean mass density. \textbf{e.} Water diffusion coefficient. \textbf{f.} First oxygen-oxygen coordination number.}
    \label{fig:uio66-water-allegro}
\end{figure*}

Water in nano-porous materials constitutes an extra challenging system due to the combination of a vast configurational space associated with water as a guest molecule and the potential for the material to catalyze chemical transformations, such as proton hopping. In Figure~\ref{fig:uio66-water-allegro}, we report a benchmark of CAGO-based active learning with DC-R$^2$SCAN for water adsorption inside UiO-66, a zirconium-based metal-organic framework (MOF) with a very large surface area as well as high thermal stability.\cite{cavka2008new} We start the training from 11 structures with different water content (see Supplementary Information). For all benchmarked properties, we see a systematic convergence with training set data just after $\sim$250 training set structures. This is a higher number than for liquid water, but this system also involves two additional chemical species with a higher level of chemical variability. To put this into context, a similar study of zeolite-OSADA pair with adversarial active learning started from a total of 17\,492 structures and achieved a 97\% stability rate after adding 573 adversarial structures.\cite{schwalbe-koda_differentiable_2021} In another impressive study, a training set of 400\,000 energies and forces from adversarial active learning, using gas-phase calculations, was used to develop an MLIP for silica with reactive water.\cite{roy2024learning}

\section{Discussion}
In this paper, we presented the CAGO algorithm to systematically generate adversarial molecular structures with user-assigned MLIP prediction errors. Through our benchmarks of the different hyperparameters of CAGO, we establish suitable minimums. In particular, we found it important to use many models, about 10 models in our case, to obtain a good error calibration, while the CAGO algorithm itself works well even for small committees. For the true error to converge to the target error, moderate error targets should be used. This is in line with our error calibration analysis, where we saw that error calibration is less robust for the high-error regime, where calibration data is scarce. 

These insights enabled us to perform active learning using minimal datasets with simple active learning protocols. In particular, we learned liquid water from a single structure of water for both Allegro and DeePMD and only 11 for water adsorption in the UiO-66 metal-organic framework. Although CAGO seemingly introduces an extra step of complexity into the active learning scheme, it ultimately simplifies the rest of the workflow. With CAGO, fewer manual adjustments and tricks are needed, such as low/high-temperature sampling or reducing the size of MD steps to achieve accurate models. Moreover, we have used orders of magnitude fewer structures than is typically reported in other works on adversarial active learning.\cite{schwalbe-koda_differentiable_2021,roy2024learning} A key problem with current MLIPs is their typical dependence on large and approximate but cheap DFT datasets. CAGO-based active learning offers a promising solution by only requiring a few structures, allowing highly accurate but costly coupled-cluster and quantum Monte Carlo reference calculations to be used as training data.

\section{Methods}
\subsection{CAGO algorithm and implementation}
The CAGO algorithm works much like ordinary molecular geometry/structure optimization, but instead of minimizing the energy, CAGO minimizes the fitness function in equation~\eqref{eq:opt-problem}, to obtain a prescribed uncertainty for the MLIPs. In particular, we use the L-BFGS algorithm, as implemented in the SciPy library,\cite{virtanen_scipy_2020} to optimize equation~\eqref{eq:opt-problem} with respect to scaled coordinates and cell vectors. The gradients are computed numerically by a two-point finite difference operator. While computing the hessian analytically is possible and most likely faster for differentiable MLIP architectures, as done in ref.~\citenum{schwalbe-koda_differentiable_2021}, numerical derivatives have certain advantages. First, single-point calculations are fast with MLIPs, making the 3$N$+6+1 calculations needed for numerical gradient calculations feasible. Second, these single-point calculations can be prepared for batch-based calculations, rather than being performed one by one, which is efficient with GPUs. Third, the calculations are trivially parallelizable, making linear scaling by adding more GPUs possible. Fourth, Hessians are generally not highly optimized with PyTorch and incur significant overheads. Finally and most importantly, while single-point calculations are ubiquitous among MLIPs, Hessians are rarely available out of the box and, therefore, not amenable to general active learning pipelines, forcing a reliance on specific MLIP architectures that may not be state-of-the-art.

\subsection{Machine learning interatomic potentials}\label{sec:MLIPs}
We tested CAGO on two different MLIPs with different architectures: \emph{Allegro} and \emph{DeepMD}. 
\paragraph{Allegro} 
For Allegro models,\cite{musaelian_learning_2023} we used equivariant E(3) products up to $L_\text{max} = 2$ in the tensor layers, with two interaction layers, including three tensor product layers of 128 neurons each. Before the interaction layer, a feature layer with 16 input features was used. Following the interaction layer block, we used three latent layers, each containing 128 neurons. A polynomial cutoff of 6~Å was used with a trainable Bessel basis set of 8 basis functions to form the feature descriptors. We used a radial cutoff of 6~Å for building the neighbor lists. Our loss function used forces, energy, and stresses, with coefficients 1, 5, and 100, respectively. These settings are similar to those previously used to generate water MLIPs with Allegro.\cite{maxson2024transferable}

\paragraph{DeePMD}
For DeePMD models,\cite{zeng_deepmd-kit_2023} we used the se\_e2\_a architecture and 25, 50, 100 neurons for the hidden embedding layer, with the submatrix of the embedding matrix using 16 neurons. Similarly to Allegro, a distance cutoff of 6~Å was used, but with a smoothing region of 0.5~Å. The potential is represented by a fully connected deep neural network with three layers of 240 neurons each. These settings have also been used before in the study of different water phases.\cite{zhai_short_2023,bore_realistic_2023} 

Input files detailing the setup of MLIPs are available in the data repository associated with this manuscript.

\subsection{Reference calculations for training and test set data} \label{sec:ref}
\paragraph*{DNN@MB-pol} For benchmarking CAGO in \ref{sec:res-cali}, we ran reference caclulations using the DNN@MB-pol model developed in refs.~\citenum{bore_realistic_2023,sciortino2025constraints} based on the MB-pol water model,\cite{babin2013development,babin2014development} using the se\_e2\_a DeePMD MLIP architecture.\cite{zeng_deepmd-kit_2023} 

\paragraph*{DC-R$^2$SCAN} To benchmark CAGO as part of active learning, we employed the DC-R$^2$SCAN method. To calculate the DC-R$^2$SCAN reference energies, forces, and virials for the MLIPs, we used the recent implementation for density-corrected DFT in CP2K,\cite{belleflamme_radicals_2023} which has in a series of papers demonstrated great accuracy with benchmarks against coupled-cluster levels of theory.\cite{dasgupta_elevating_2021,dasgupta_nuclear_2024} The atomic core electrons were described using Goedecker–Teter–Hutter (GTH) pseudopotentials, while valence electron molecular orbitals were expanded in triple-zeta double-polarized basis sets (TZV2P) optimized for the SCAN functional. The kinetic energy cutoff for the plane-wave expansion of the density was set to 2500~Ry, which was required to get a numerical accuracy of $\sim$2~meV/\AA\, and $\sim$0.05 kBar for forces and stress-tensor trace, respectively. The truncated Coulomb operator with a cutoff radius between 5.0 and 5.5 {\AA} was used depending on the system, corresponding to approximately half the length of the smallest edge of the simulation cell. To overcome the expense of the Hartree-Fock calculation, the ADMM approximation was used with an optimized valence basis set (BASIS\_ADMM\_UZH). The Schwarz integral screening threshold was set to $10^{-6}$ atomic units, starting from a converged PBE calculation. All the DC-R$^2$SCAN calculations in this work were carried out with CP2K, with associated inputs and outputs stored in the data repository associated with this manuscript.  

\subsection{CAGO-based active learning procedure and implementation}
The CAGO-based active learning procedure is an iterative process depicted in Figure~\ref{fig:al}. Our implementation begins with an initial set of reference data, reserving 10~\% of this data for a test set. The remaining 90~\% reference data is partitioned into an 80~\% training set and a 10~\% validation set through bootstrapping for each machine learning interatomic potential (MLIP) committee member, as described in ref.~\citenum{palmer_calibration_2022}. After training the committee of MLIPs, their uncertainty is calibrated using the test set. Subsequently, MD simulations are conducted with the MLIP committee. From MD simulations, a subset of structures are randomly sampled to create adversarial structures with CAGO. To reduce the risk for failed reference calculations, a filter is applied to these structures, such as an upper threshold for the mean force magnitude of the MLIP committee. Reference calculations are then performed on the filtered adversarial structures before undergoing a final filter, checking for convergence and avoiding unphysical structures dominated by sterics before incorporating them into the reference data set, completing one iteration of the active learning cycle. This entire active learning loop is implemented in the Hylleraas Software Platform,\cite{hsp} which interfaces with various MLIPs and quantum chemistry softwares, enabling the use of heterogeneous computing environments on high-performance computer clusters.

\subsection{Computational procedures}\label{sec:sim-details}
\subsubsection{Calibrated adversarial geometry optimization to target error}\label{sec:cago-benchmark}
For benchmarking the CAGO algorithm in section~\ref{sec:res-cali}, we targeted the DNN@MB-pol water model as specified in Methods~\ref{sec:ref}. First, we trained a committee of 20 DeePMD MLIPs on a training set of MD trajectories. Second, using these 20 models and a separate test set, we performed uncertainty calibration in accordance with equation~\eqref{eq:log-likelihood}, the results of which are reported in Figure~\ref{fig:adv-illu}. Third, sampling 40 structures from the test set, we perform CAGO with various hyperparameters, such as different error targets and committee sizes. The errors we report are root mean square errors of individual models against DNN@MB-pol. For more details on the training and test set, we refer to the Supplementary Information. 

\subsubsection{Learning liquid water from a single structure}\label{sec:stability}
For the active learning workflow, we used committees with 12 MLIPs to learn liquid water from a single structure of 64 water molecules, with details on initial training sets reported in Supplementary Information. In each of the first two active learning iterations with CAGO, we sampled five structures from the training set and optimized the structures with CAGO using random subsets of 3 committee members.  For all subsequent iterations, MD sampling with LAMMPS~\cite{thompson_lammps_2022} was performed before CAGO (see Section~\ref{sec:sim-details}), and in the case of pure water, 20 samples were extracted and taken for CAGO considering all 12 committee members. Due to the low number of structures in the first iterations of active learning, the MLIPs tend to be unstable; we, therefore, performed $NVT$ simulations at 300~K to generate new structures for CAGO for iterations 3 and 4. These simulations were run for 50~ps using the velocity-Verlet integrator with a 0.25~fs timestep and Nosé-Hoover thermostat chain with 3 thermostats and a 0.5~ps relaxation time. From iteration 5 and onwards, we performed $NPT$ simulations at 300~K and 1~bar, sampling for 100~ps, using the same settings for the thermostat part, and a Nosé-Hoover barostat chain of 3 barostats with 1~ps relaxation time. Structures derived from CAGO with maximum forces (mean by committee members) higher than 40~eV/\AA\ or maximum force from reference calculations higher than 30~eV/\AA, or minimum distance lower than 0.75~\AA, or with convergence warnings from CP2K, were filtered out. Note that the DeePMD MLIPs used slightly varying settings for the training throughout the active learning loop to avoid overfitting. In the first two iterations, we trained the models for 50\,000 steps to avoid overfitting with small training sets. As more structures were added, we increased the number of training steps to 150\,000 for the next two active learning iterations, and from iteration 5 and onwards, we used 400\,000 training steps. For active learning with maximum uncertainty and elevated temperature, MD-sampled structures were used for all iterations. 500~K was used for elevated temperature active learning. For maximum uncertainty, we picked structures based on the highest force uncertainty on any atom in the structure. For elevated temperature active learning, we sampled structures at random. In both cases we filter high-force structures as in the CAGO-based active learning.

For the benchmark of MLIPs along active learning iterations in Figure~\ref{fig:water-deepmd-allegro}, we ran 3 independent simulations starting from different initial conditions in a simulation box with 256 water molecules. We first performed two thermalizations for each system, one in $NVT$ (25~ps) and another in the $NPT$ (100~ps) ensemble, using the Berendsen thermostat and barostat. Finally, we ran a 2~ns $NPT$ simulation using the Nosé-Hoover thermostat and barostat. These simulations used a 0.5~fs timestep, and all other settings were the same as the ones described above for the sampling phase of active learning.

We use two criteria to define the stability of the MLIPs during MD simulations. First, we check every 100 timesteps that the minimal distance between atoms in the system is never below 0.4~\AA. Second, we monitor the system density, verifying that it does not diverge to values lower than 0.25 times or higher than 2 times the initial density. Furthermore, we ran multiple replicas, considering the MLIP to be stable only if all trajectories of the replicas meet our stability criteria.

\subsubsection{Learning water-adsorption of a metal-organic framework}
The water adsorption in the UiO-66 simulations was performed using similar workflow settings as for liquid water in Methods~\ref{sec:stability}, with a few modifications as follows. In particular, for this active learning workflow, we considered multiple systems, including liquid water and 10 structures of UiO-66 with differing water content (see Supplementary Information). Therefore, in this workflow, we started from an initial reference data set of 11 structures and extended the reference data set for each individual system. In each of the first two active learning iterations, we sampled five structures from the training set and optimized the structures with CAGO and random subsets of 3 committee members for each of the systems. For the rest of the iterations, we performed $NPT$ MD sampling for each system with all committee members, followed by CAGO on five structures with four committee members for each of the 11 systems. We used a timestep of 0.5~fs and ran the simulations for 100~ps during the active learning sampling phase. The benchmark of MLIPs along active learning iterations in Figure~\ref{fig:uio66-water-allegro} were performed using the system presented in Figure~\ref{fig:uio66-water-allegro}a, which is UiO-66 with a water weight percentage of 45, with the same MD protocol as in Methods~\ref{sec:stability}.

\section{Data availability}
The dataset and scripts used to produce the data in this study are publicly available via the Norwegian {\it National e-Infrastructure for Research Data} (NIRD) at \url{https://doi.org/10.11582/2025.00018}.

\section{Code availability}
The code implementing the active learning loop using CAGO is publicly available at GitLab: \url{https://gitlab.com/hylleraasplatform/hyal}.

\section{Acknowledgments}
We would like to thank the anonymous reviewer for his/her suggestion of adding a comparison with standard active learning based on picking maximum uncertainty structures from MD, and random structures from elevated temperature MD. This work was supported by the Research Council of Norway through the Centre of Excellence Hylleraas Centre for Quantum Molecular Sciences (Grant 262695) and the Young Researcher Talent grants 344993 and 354100. We acknowledge the EuroHPC Joint Undertaking for awarding this project access to the EuroHPC supercomputer LUMI, hosted by CSC (Finland) and the LUMI consortium through a EuroHPC Regular Access call (Grants EHPC-REG-2023R02-088, EHPC-REG-2023R03-146). Support was also received from the Centre for Advanced Study in Oslo, Norway, which funded and hosted the SLB Young CAS Fellow research project during the academic year of 23/24 and 24/25. Part of the simulations were performed on resources provided by Sigma2 — the Norwegian National Infrastructure for High-Performance Computing and Data Storage (grant numbers NN4654K and NS4654K).

\bibliographystyle{apsrev4-1}
\bibliography{mybib}

\begin{thebibliography}{42}%
\makeatletter
\providecommand \@ifxundefined [1]{%
 \@ifx{#1\undefined}
}%
\providecommand \@ifnum [1]{%
 \ifnum #1\expandafter \@firstoftwo
 \else \expandafter \@secondoftwo
 \fi
}%
\providecommand \@ifx [1]{%
 \ifx #1\expandafter \@firstoftwo
 \else \expandafter \@secondoftwo
 \fi
}%
\providecommand \natexlab [1]{#1}%
\providecommand \enquote  [1]{``#1''}%
\providecommand \bibnamefont  [1]{#1}%
\providecommand \bibfnamefont [1]{#1}%
\providecommand \citenamefont [1]{#1}%
\providecommand \href@noop [0]{\@secondoftwo}%
\providecommand \href [0]{\begingroup \@sanitize@url \@href}%
\providecommand \@href[1]{\@@startlink{#1}\@@href}%
\providecommand \@@href[1]{\endgroup#1\@@endlink}%
\providecommand \@sanitize@url [0]{\catcode `\\12\catcode `\$12\catcode
  `\&12\catcode `\#12\catcode `\^12\catcode `\_12\catcode `\%12\relax}%
\providecommand \@@startlink[1]{}%
\providecommand \@@endlink[0]{}%
\providecommand \url  [0]{\begingroup\@sanitize@url \@url }%
\providecommand \@url [1]{\endgroup\@href {#1}{\urlprefix }}%
\providecommand \urlprefix  [0]{URL }%
\providecommand \Eprint [0]{\href }%
\providecommand \doibase [0]{http://dx.doi.org/}%
\providecommand \selectlanguage [0]{\@gobble}%
\providecommand \bibinfo  [0]{\@secondoftwo}%
\providecommand \bibfield  [0]{\@secondoftwo}%
\providecommand \translation [1]{[#1]}%
\providecommand \BibitemOpen [0]{}%
\providecommand \bibitemStop [0]{}%
\providecommand \bibitemNoStop [0]{.\EOS\space}%
\providecommand \EOS [0]{\spacefactor3000\relax}%
\providecommand \BibitemShut  [1]{\csname bibitem#1\endcsname}%
\let\auto@bib@innerbib\@empty
\bibitem [{\citenamefont {Behler}\ and\ \citenamefont
  {Parrinello}(2007)}]{behler_generalized_2007}%
  \BibitemOpen
  \bibfield  {author} {\bibinfo {author} {\bibfnamefont {J.}~\bibnamefont
  {Behler}}\ and\ \bibinfo {author} {\bibfnamefont {M.}~\bibnamefont
  {Parrinello}},\ }\href {\doibase 10.1103/PhysRevLett.98.146401} {\bibfield
  {journal} {\bibinfo  {journal} {Physical Review Letters}\ }\textbf {\bibinfo
  {volume} {98}},\ \bibinfo {pages} {146401} (\bibinfo {year}
  {2007})}\BibitemShut {NoStop}%
\bibitem [{\citenamefont {Batzner}\ \emph {et~al.}(2022)\citenamefont
  {Batzner}, \citenamefont {Musaelian}, \citenamefont {Sun}, \citenamefont
  {Geiger}, \citenamefont {Mailoa}, \citenamefont {Kornbluth}, \citenamefont
  {Molinari}, \citenamefont {Smidt},\ and\ \citenamefont
  {Kozinsky}}]{batzner_e3-equivariant_2022}%
  \BibitemOpen
  \bibfield  {author} {\bibinfo {author} {\bibfnamefont {S.}~\bibnamefont
  {Batzner}}, \bibinfo {author} {\bibfnamefont {A.}~\bibnamefont {Musaelian}},
  \bibinfo {author} {\bibfnamefont {L.}~\bibnamefont {Sun}}, \bibinfo {author}
  {\bibfnamefont {M.}~\bibnamefont {Geiger}}, \bibinfo {author} {\bibfnamefont
  {J.~P.}\ \bibnamefont {Mailoa}}, \bibinfo {author} {\bibfnamefont
  {M.}~\bibnamefont {Kornbluth}}, \bibinfo {author} {\bibfnamefont
  {N.}~\bibnamefont {Molinari}}, \bibinfo {author} {\bibfnamefont {T.~E.}\
  \bibnamefont {Smidt}}, \ and\ \bibinfo {author} {\bibfnamefont
  {B.}~\bibnamefont {Kozinsky}},\ }\href {\doibase 10.1038/s41467-022-29939-5}
  {\bibfield  {journal} {\bibinfo  {journal} {Nature Communications}\ }\textbf
  {\bibinfo {volume} {13}},\ \bibinfo {pages} {2453} (\bibinfo {year}
  {2022})}\BibitemShut {NoStop}%
\bibitem [{\citenamefont {Musaelian}\ \emph {et~al.}(2023)\citenamefont
  {Musaelian}, \citenamefont {Batzner}, \citenamefont {Johansson},
  \citenamefont {Sun}, \citenamefont {Owen}, \citenamefont {Kornbluth},\ and\
  \citenamefont {Kozinsky}}]{musaelian_learning_2023}%
  \BibitemOpen
  \bibfield  {author} {\bibinfo {author} {\bibfnamefont {A.}~\bibnamefont
  {Musaelian}}, \bibinfo {author} {\bibfnamefont {S.}~\bibnamefont {Batzner}},
  \bibinfo {author} {\bibfnamefont {A.}~\bibnamefont {Johansson}}, \bibinfo
  {author} {\bibfnamefont {L.}~\bibnamefont {Sun}}, \bibinfo {author}
  {\bibfnamefont {C.~J.}\ \bibnamefont {Owen}}, \bibinfo {author}
  {\bibfnamefont {M.}~\bibnamefont {Kornbluth}}, \ and\ \bibinfo {author}
  {\bibfnamefont {B.}~\bibnamefont {Kozinsky}},\ }\href {\doibase
  10.1038/s41467-023-36329-y} {\bibfield  {journal} {\bibinfo  {journal}
  {Nature Communications}\ }\textbf {\bibinfo {volume} {14}},\ \bibinfo {pages}
  {579} (\bibinfo {year} {2023})}\BibitemShut {NoStop}%
\bibitem [{\citenamefont {Batatia}\ \emph {et~al.}(2022)\citenamefont
  {Batatia}, \citenamefont {Kovacs}, \citenamefont {Simm}, \citenamefont
  {Ortner},\ and\ \citenamefont {Csanyi}}]{NEURIPS2022_4a36c3c5}%
  \BibitemOpen
  \bibfield  {author} {\bibinfo {author} {\bibfnamefont {I.}~\bibnamefont
  {Batatia}}, \bibinfo {author} {\bibfnamefont {D.~P.}\ \bibnamefont {Kovacs}},
  \bibinfo {author} {\bibfnamefont {G.}~\bibnamefont {Simm}}, \bibinfo {author}
  {\bibfnamefont {C.}~\bibnamefont {Ortner}}, \ and\ \bibinfo {author}
  {\bibfnamefont {G.}~\bibnamefont {Csanyi}},\ }in\ \href
  {https://proceedings.neurips.cc/paper_files/paper/2022/file/4a36c3c51af11ed9f34615b81edb5bbc-Paper-Conference.pdf}
  {\emph {\bibinfo {booktitle} {Advances in Neural Information Processing
  Systems}}},\ Vol.~\bibinfo {volume} {35},\ \bibinfo {editor} {edited by\
  \bibinfo {editor} {\bibfnamefont {S.}~\bibnamefont {Koyejo}}, \bibinfo
  {editor} {\bibfnamefont {S.}~\bibnamefont {Mohamed}}, \bibinfo {editor}
  {\bibfnamefont {A.}~\bibnamefont {Agarwal}}, \bibinfo {editor} {\bibfnamefont
  {D.}~\bibnamefont {Belgrave}}, \bibinfo {editor} {\bibfnamefont
  {K.}~\bibnamefont {Cho}}, \ and\ \bibinfo {editor} {\bibfnamefont
  {A.}~\bibnamefont {Oh}}}\ (\bibinfo  {publisher} {Curran Associates, Inc.},\
  \bibinfo {year} {2022})\ pp.\ \bibinfo {pages} {11423--11436}\BibitemShut
  {NoStop}%
\bibitem [{\citenamefont {Zhai}\ \emph {et~al.}(2023)\citenamefont {Zhai},
  \citenamefont {Caruso}, \citenamefont {Bore}, \citenamefont {Luo},\ and\
  \citenamefont {Paesani}}]{zhai_short_2023}%
  \BibitemOpen
  \bibfield  {author} {\bibinfo {author} {\bibfnamefont {Y.}~\bibnamefont
  {Zhai}}, \bibinfo {author} {\bibfnamefont {A.}~\bibnamefont {Caruso}},
  \bibinfo {author} {\bibfnamefont {S.~L.}\ \bibnamefont {Bore}}, \bibinfo
  {author} {\bibfnamefont {Z.}~\bibnamefont {Luo}}, \ and\ \bibinfo {author}
  {\bibfnamefont {F.}~\bibnamefont {Paesani}},\ }\href {\doibase
  10.1063/5.0142843} {\bibfield  {journal} {\bibinfo  {journal} {The Journal of
  Chemical Physics}\ }\textbf {\bibinfo {volume} {158}},\ \bibinfo {pages}
  {084111} (\bibinfo {year} {2023})}\BibitemShut {NoStop}%
\bibitem [{\citenamefont {Fu}\ \emph {et~al.}(2023)\citenamefont {Fu},
  \citenamefont {Wu}, \citenamefont {Wang}, \citenamefont {Xie}, \citenamefont
  {Keten}, \citenamefont {Gomez-Bombarelli},\ and\ \citenamefont
  {Jaakkola}}]{fu2023forces}%
  \BibitemOpen
  \bibfield  {author} {\bibinfo {author} {\bibfnamefont {X.}~\bibnamefont
  {Fu}}, \bibinfo {author} {\bibfnamefont {Z.}~\bibnamefont {Wu}}, \bibinfo
  {author} {\bibfnamefont {W.}~\bibnamefont {Wang}}, \bibinfo {author}
  {\bibfnamefont {T.}~\bibnamefont {Xie}}, \bibinfo {author} {\bibfnamefont
  {S.}~\bibnamefont {Keten}}, \bibinfo {author} {\bibfnamefont
  {R.}~\bibnamefont {Gomez-Bombarelli}}, \ and\ \bibinfo {author}
  {\bibfnamefont {T.}~\bibnamefont {Jaakkola}},\ }\href
  {https://openreview.net/forum?id=A8pqQipwkt} {\bibfield  {journal} {\bibinfo
  {journal} {Transactions on Machine Learning Research}\ } (\bibinfo {year}
  {2023})}\BibitemShut {NoStop}%
\bibitem [{\citenamefont {Shimodaira}(2000)}]{shimodaira_improving_2000}%
  \BibitemOpen
  \bibfield  {author} {\bibinfo {author} {\bibfnamefont {H.}~\bibnamefont
  {Shimodaira}},\ }\href {\doibase 10.1016/S0378-3758(00)00115-4} {\bibfield
  {journal} {\bibinfo  {journal} {Journal of Statistical Planning and
  Inference}\ }\textbf {\bibinfo {volume} {90}},\ \bibinfo {pages} {227}
  (\bibinfo {year} {2000})}\BibitemShut {NoStop}%
\bibitem [{\citenamefont {Yang}\ \emph {et~al.}(2024)\citenamefont {Yang},
  \citenamefont {Zhang}, \citenamefont {Ranasinghe}, \citenamefont {Isayev},\
  and\ \citenamefont {Roitberg}}]{yang2024machine}%
  \BibitemOpen
  \bibfield  {author} {\bibinfo {author} {\bibfnamefont {Y.}~\bibnamefont
  {Yang}}, \bibinfo {author} {\bibfnamefont {S.}~\bibnamefont {Zhang}},
  \bibinfo {author} {\bibfnamefont {K.~D.}\ \bibnamefont {Ranasinghe}},
  \bibinfo {author} {\bibfnamefont {O.}~\bibnamefont {Isayev}}, \ and\ \bibinfo
  {author} {\bibfnamefont {A.~E.}\ \bibnamefont {Roitberg}},\ }\href
  {https://doi.org/10.1146/annurev-physchem-062123-024417} {\bibfield
  {journal} {\bibinfo  {journal} {Annual Review of Physical Chemistry}\
  }\textbf {\bibinfo {volume} {75}},\ \bibinfo {pages} {371} (\bibinfo {year}
  {2024})}\BibitemShut {NoStop}%
\bibitem [{\citenamefont {Vandermause}\ \emph {et~al.}(2020)\citenamefont
  {Vandermause}, \citenamefont {Torrisi}, \citenamefont {Batzner},
  \citenamefont {Xie}, \citenamefont {Sun}, \citenamefont {Kolpak},\ and\
  \citenamefont {Kozinsky}}]{vandermause_--fly_2020}%
  \BibitemOpen
  \bibfield  {author} {\bibinfo {author} {\bibfnamefont {J.}~\bibnamefont
  {Vandermause}}, \bibinfo {author} {\bibfnamefont {S.~B.}\ \bibnamefont
  {Torrisi}}, \bibinfo {author} {\bibfnamefont {S.}~\bibnamefont {Batzner}},
  \bibinfo {author} {\bibfnamefont {Y.}~\bibnamefont {Xie}}, \bibinfo {author}
  {\bibfnamefont {L.}~\bibnamefont {Sun}}, \bibinfo {author} {\bibfnamefont
  {A.~M.}\ \bibnamefont {Kolpak}}, \ and\ \bibinfo {author} {\bibfnamefont
  {B.}~\bibnamefont {Kozinsky}},\ }\href {\doibase 10.1038/s41524-020-0283-z}
  {\bibfield  {journal} {\bibinfo  {journal} {npj Computational Materials}\
  }\textbf {\bibinfo {volume} {6}},\ \bibinfo {pages} {20} (\bibinfo {year}
  {2020})}\BibitemShut {NoStop}%
\bibitem [{\citenamefont {Xie}\ \emph {et~al.}(2023)\citenamefont {Xie},
  \citenamefont {Vandermause}, \citenamefont {Ramakers}, \citenamefont
  {Protik}, \citenamefont {Johansson},\ and\ \citenamefont
  {Kozinsky}}]{xie_uncertainty-aware_2023}%
  \BibitemOpen
  \bibfield  {author} {\bibinfo {author} {\bibfnamefont {Y.}~\bibnamefont
  {Xie}}, \bibinfo {author} {\bibfnamefont {J.}~\bibnamefont {Vandermause}},
  \bibinfo {author} {\bibfnamefont {S.}~\bibnamefont {Ramakers}}, \bibinfo
  {author} {\bibfnamefont {N.~H.}\ \bibnamefont {Protik}}, \bibinfo {author}
  {\bibfnamefont {A.}~\bibnamefont {Johansson}}, \ and\ \bibinfo {author}
  {\bibfnamefont {B.}~\bibnamefont {Kozinsky}},\ }\href {\doibase
  10.1038/s41524-023-00988-8} {\bibfield  {journal} {\bibinfo  {journal} {npj
  Computational Materials}\ }\textbf {\bibinfo {volume} {9}},\ \bibinfo {pages}
  {36} (\bibinfo {year} {2023})}\BibitemShut {NoStop}%
\bibitem [{\citenamefont {Wen}\ and\ \citenamefont
  {Tadmor}(2020)}]{wen_uncertainty_2020}%
  \BibitemOpen
  \bibfield  {author} {\bibinfo {author} {\bibfnamefont {M.}~\bibnamefont
  {Wen}}\ and\ \bibinfo {author} {\bibfnamefont {E.~B.}\ \bibnamefont
  {Tadmor}},\ }\href {\doibase 10.1038/s41524-020-00390-8} {\bibfield
  {journal} {\bibinfo  {journal} {npj Computational Materials}\ }\textbf
  {\bibinfo {volume} {6}},\ \bibinfo {pages} {124} (\bibinfo {year}
  {2020})}\BibitemShut {NoStop}%
\bibitem [{\citenamefont {Krogh}\ and\ \citenamefont
  {Vedelsby}(1994)}]{NIPS1994_b8c37e33}%
  \BibitemOpen
  \bibfield  {author} {\bibinfo {author} {\bibfnamefont {A.}~\bibnamefont
  {Krogh}}\ and\ \bibinfo {author} {\bibfnamefont {J.}~\bibnamefont
  {Vedelsby}},\ }in\ \href
  {https://proceedings.neurips.cc/paper_files/paper/1994/file/b8c37e33defde51cf91e1e03e51657da-Paper.pdf}
  {\emph {\bibinfo {booktitle} {Advances in Neural Information Processing
  Systems}}},\ Vol.~\bibinfo {volume} {7},\ \bibinfo {editor} {edited by\
  \bibinfo {editor} {\bibfnamefont {G.}~\bibnamefont {Tesauro}}, \bibinfo
  {editor} {\bibfnamefont {D.}~\bibnamefont {Touretzky}}, \ and\ \bibinfo
  {editor} {\bibfnamefont {T.}~\bibnamefont {Leen}}}\ (\bibinfo  {publisher}
  {MIT Press},\ \bibinfo {year} {1994})\BibitemShut {NoStop}%
\bibitem [{\citenamefont {Smith}\ \emph
  {et~al.}(2018{\natexlab{a}})\citenamefont {Smith}, \citenamefont {Nebgen},
  \citenamefont {Lubbers}, \citenamefont {Isayev},\ and\ \citenamefont
  {Roitberg}}]{smith_less_2018}%
  \BibitemOpen
  \bibfield  {author} {\bibinfo {author} {\bibfnamefont {J.~S.}\ \bibnamefont
  {Smith}}, \bibinfo {author} {\bibfnamefont {B.}~\bibnamefont {Nebgen}},
  \bibinfo {author} {\bibfnamefont {N.}~\bibnamefont {Lubbers}}, \bibinfo
  {author} {\bibfnamefont {O.}~\bibnamefont {Isayev}}, \ and\ \bibinfo {author}
  {\bibfnamefont {A.~E.}\ \bibnamefont {Roitberg}},\ }\href {\doibase
  10.1063/1.5023802} {\bibfield  {journal} {\bibinfo  {journal} {The Journal of
  Chemical Physics}\ }\textbf {\bibinfo {volume} {148}},\ \bibinfo {pages}
  {241733} (\bibinfo {year} {2018}{\natexlab{a}})}\BibitemShut {NoStop}%
\bibitem [{\citenamefont {Zhang}\ \emph {et~al.}(2020)\citenamefont {Zhang},
  \citenamefont {Wang}, \citenamefont {Chen}, \citenamefont {Zeng},
  \citenamefont {Zhang}, \citenamefont {Wang},\ and\ \citenamefont
  {E}}]{zhang_dp-gen_2020}%
  \BibitemOpen
  \bibfield  {author} {\bibinfo {author} {\bibfnamefont {Y.}~\bibnamefont
  {Zhang}}, \bibinfo {author} {\bibfnamefont {H.}~\bibnamefont {Wang}},
  \bibinfo {author} {\bibfnamefont {W.}~\bibnamefont {Chen}}, \bibinfo {author}
  {\bibfnamefont {J.}~\bibnamefont {Zeng}}, \bibinfo {author} {\bibfnamefont
  {L.}~\bibnamefont {Zhang}}, \bibinfo {author} {\bibfnamefont
  {H.}~\bibnamefont {Wang}}, \ and\ \bibinfo {author} {\bibfnamefont
  {W.}~\bibnamefont {E}},\ }\href {\doibase 10.1016/j.cpc.2020.107206}
  {\bibfield  {journal} {\bibinfo  {journal} {Computer Physics Communications}\
  }\textbf {\bibinfo {volume} {253}},\ \bibinfo {pages} {107206} (\bibinfo
  {year} {2020})}\BibitemShut {NoStop}%
\bibitem [{\citenamefont {Schran}\ \emph {et~al.}(2021)\citenamefont {Schran},
  \citenamefont {Thiemann}, \citenamefont {Rowe}, \citenamefont {Müller},
  \citenamefont {Marsalek},\ and\ \citenamefont
  {Michaelides}}]{schran_machine_2021}%
  \BibitemOpen
  \bibfield  {author} {\bibinfo {author} {\bibfnamefont {C.}~\bibnamefont
  {Schran}}, \bibinfo {author} {\bibfnamefont {F.~L.}\ \bibnamefont
  {Thiemann}}, \bibinfo {author} {\bibfnamefont {P.}~\bibnamefont {Rowe}},
  \bibinfo {author} {\bibfnamefont {E.~A.}\ \bibnamefont {Müller}}, \bibinfo
  {author} {\bibfnamefont {O.}~\bibnamefont {Marsalek}}, \ and\ \bibinfo
  {author} {\bibfnamefont {A.}~\bibnamefont {Michaelides}},\ }\href {\doibase
  10.1073/pnas.2110077118} {\bibfield  {journal} {\bibinfo  {journal}
  {Proceedings of the National Academy of Sciences}\ }\textbf {\bibinfo
  {volume} {118}},\ \bibinfo {pages} {e2110077118} (\bibinfo {year}
  {2021})}\BibitemShut {NoStop}%
\bibitem [{\citenamefont {Vandenhaute}\ \emph {et~al.}(2023)\citenamefont
  {Vandenhaute}, \citenamefont {Cools-Ceuppens}, \citenamefont {DeKeyser},
  \citenamefont {Verstraelen},\ and\ \citenamefont
  {Van~Speybroeck}}]{vandenhaute_machine_2023}%
  \BibitemOpen
  \bibfield  {author} {\bibinfo {author} {\bibfnamefont {S.}~\bibnamefont
  {Vandenhaute}}, \bibinfo {author} {\bibfnamefont {M.}~\bibnamefont
  {Cools-Ceuppens}}, \bibinfo {author} {\bibfnamefont {S.}~\bibnamefont
  {DeKeyser}}, \bibinfo {author} {\bibfnamefont {T.}~\bibnamefont
  {Verstraelen}}, \ and\ \bibinfo {author} {\bibfnamefont {V.}~\bibnamefont
  {Van~Speybroeck}},\ }\href {\doibase 10.1038/s41524-023-00969-x} {\bibfield
  {journal} {\bibinfo  {journal} {npj Computational Materials}\ }\textbf
  {\bibinfo {volume} {9}},\ \bibinfo {pages} {19} (\bibinfo {year}
  {2023})}\BibitemShut {NoStop}%
\bibitem [{\citenamefont {Yang}\ \emph {et~al.}(2022)\citenamefont {Yang},
  \citenamefont {Bonati}, \citenamefont {Polino},\ and\ \citenamefont
  {Parrinello}}]{yang_using_2022}%
  \BibitemOpen
  \bibfield  {author} {\bibinfo {author} {\bibfnamefont {M.}~\bibnamefont
  {Yang}}, \bibinfo {author} {\bibfnamefont {L.}~\bibnamefont {Bonati}},
  \bibinfo {author} {\bibfnamefont {D.}~\bibnamefont {Polino}}, \ and\ \bibinfo
  {author} {\bibfnamefont {M.}~\bibnamefont {Parrinello}},\ }\href {\doibase
  10.1016/j.cattod.2021.03.018} {\bibfield  {journal} {\bibinfo  {journal}
  {Catalysis Today}\ }\textbf {\bibinfo {volume} {387}},\ \bibinfo {pages}
  {143} (\bibinfo {year} {2022})}\BibitemShut {NoStop}%
\bibitem [{\citenamefont {Smith}\ \emph
  {et~al.}(2018{\natexlab{b}})\citenamefont {Smith}, \citenamefont {Nebgen},
  \citenamefont {Lubbers}, \citenamefont {Isayev},\ and\ \citenamefont
  {Roitberg}}]{smith2018less}%
  \BibitemOpen
  \bibfield  {author} {\bibinfo {author} {\bibfnamefont {J.~S.}\ \bibnamefont
  {Smith}}, \bibinfo {author} {\bibfnamefont {B.}~\bibnamefont {Nebgen}},
  \bibinfo {author} {\bibfnamefont {N.}~\bibnamefont {Lubbers}}, \bibinfo
  {author} {\bibfnamefont {O.}~\bibnamefont {Isayev}}, \ and\ \bibinfo {author}
  {\bibfnamefont {A.~E.}\ \bibnamefont {Roitberg}},\ }\href
  {http://dx.doi.org/10.1063/1.5023802} {\bibfield  {journal} {\bibinfo
  {journal} {The Journal of Chemical Physics}\ }\textbf {\bibinfo {volume}
  {148}} (\bibinfo {year} {2018}{\natexlab{b}})}\BibitemShut {NoStop}%
\bibitem [{\citenamefont {Zeng}\ \emph {et~al.}(2020)\citenamefont {Zeng},
  \citenamefont {Cao}, \citenamefont {Xu}, \citenamefont {Zhu},\ and\
  \citenamefont {Zhang}}]{zeng2020complex}%
  \BibitemOpen
  \bibfield  {author} {\bibinfo {author} {\bibfnamefont {J.}~\bibnamefont
  {Zeng}}, \bibinfo {author} {\bibfnamefont {L.}~\bibnamefont {Cao}}, \bibinfo
  {author} {\bibfnamefont {M.}~\bibnamefont {Xu}}, \bibinfo {author}
  {\bibfnamefont {T.}~\bibnamefont {Zhu}}, \ and\ \bibinfo {author}
  {\bibfnamefont {J.~Z.}\ \bibnamefont {Zhang}},\ }\href
  {https://doi.org/10.1038/s41467-020-19497-z} {\bibfield  {journal} {\bibinfo
  {journal} {Nature communications}\ }\textbf {\bibinfo {volume} {11}},\
  \bibinfo {pages} {5713} (\bibinfo {year} {2020})}\BibitemShut {NoStop}%
\bibitem [{\citenamefont {Goodfellow}(2016)}]{goodfellow2016deep}%
  \BibitemOpen
  \bibfield  {author} {\bibinfo {author} {\bibfnamefont {I.}~\bibnamefont
  {Goodfellow}},\ }\href@noop {} {\emph {\bibinfo {title} {Deep learning}}},\
  Vol.\ \bibinfo {volume} {196}\ (\bibinfo  {publisher} {MIT press},\ \bibinfo
  {year} {2016})\BibitemShut {NoStop}%
\bibitem [{\citenamefont {Cubuk}\ and\ \citenamefont
  {Schoenholz}(2020)}]{cubuk_adversarial_nodate}%
  \BibitemOpen
  \bibfield  {author} {\bibinfo {author} {\bibfnamefont {E.~D.}\ \bibnamefont
  {Cubuk}}\ and\ \bibinfo {author} {\bibfnamefont {S.~S.}\ \bibnamefont
  {Schoenholz}},\ }\href {https://ml4physicalsciences.github.io/2020/}
  {\bibfield  {journal} {\bibinfo  {journal} {3rd NeurIPS workshop on Machine
  Learning and the Physical Sciences}\ } (\bibinfo {year} {2020})}\BibitemShut
  {NoStop}%
\bibitem [{\citenamefont {Kulichenko}\ \emph {et~al.}(2023)\citenamefont
  {Kulichenko}, \citenamefont {Barros}, \citenamefont {Lubbers}, \citenamefont
  {Li}, \citenamefont {Messerly}, \citenamefont {Tretiak}, \citenamefont
  {Smith},\ and\ \citenamefont {Nebgen}}]{kulichenko_uncertainty-driven_2023}%
  \BibitemOpen
  \bibfield  {author} {\bibinfo {author} {\bibfnamefont {M.}~\bibnamefont
  {Kulichenko}}, \bibinfo {author} {\bibfnamefont {K.}~\bibnamefont {Barros}},
  \bibinfo {author} {\bibfnamefont {N.}~\bibnamefont {Lubbers}}, \bibinfo
  {author} {\bibfnamefont {Y.~W.}\ \bibnamefont {Li}}, \bibinfo {author}
  {\bibfnamefont {R.}~\bibnamefont {Messerly}}, \bibinfo {author}
  {\bibfnamefont {S.}~\bibnamefont {Tretiak}}, \bibinfo {author} {\bibfnamefont
  {J.~S.}\ \bibnamefont {Smith}}, \ and\ \bibinfo {author} {\bibfnamefont
  {B.}~\bibnamefont {Nebgen}},\ }\href {\doibase 10.1038/s43588-023-00406-5}
  {\bibfield  {journal} {\bibinfo  {journal} {Nature Computational Science}\
  }\textbf {\bibinfo {volume} {3}},\ \bibinfo {pages} {230} (\bibinfo {year}
  {2023})}\BibitemShut {NoStop}%
\bibitem [{\citenamefont {Schwalbe-Koda}\ \emph {et~al.}(2021)\citenamefont
  {Schwalbe-Koda}, \citenamefont {Tan},\ and\ \citenamefont
  {Gómez-Bombarelli}}]{schwalbe-koda_differentiable_2021}%
  \BibitemOpen
  \bibfield  {author} {\bibinfo {author} {\bibfnamefont {D.}~\bibnamefont
  {Schwalbe-Koda}}, \bibinfo {author} {\bibfnamefont {A.~R.}\ \bibnamefont
  {Tan}}, \ and\ \bibinfo {author} {\bibfnamefont {R.}~\bibnamefont
  {Gómez-Bombarelli}},\ }\href {\doibase 10.1038/s41467-021-25342-8}
  {\bibfield  {journal} {\bibinfo  {journal} {Nature Communications}\ }\textbf
  {\bibinfo {volume} {12}},\ \bibinfo {pages} {5104} (\bibinfo {year}
  {2021})}\BibitemShut {NoStop}%
\bibitem [{\citenamefont {Roy}\ \emph {et~al.}(2024)\citenamefont {Roy},
  \citenamefont {D{\"u}rholt}, \citenamefont {Asche}, \citenamefont {Zipoli},\
  and\ \citenamefont {G{\'o}mez-Bombarelli}}]{roy2024learning}%
  \BibitemOpen
  \bibfield  {author} {\bibinfo {author} {\bibfnamefont {S.}~\bibnamefont
  {Roy}}, \bibinfo {author} {\bibfnamefont {J.~P.}\ \bibnamefont
  {D{\"u}rholt}}, \bibinfo {author} {\bibfnamefont {T.~S.}\ \bibnamefont
  {Asche}}, \bibinfo {author} {\bibfnamefont {F.}~\bibnamefont {Zipoli}}, \
  and\ \bibinfo {author} {\bibfnamefont {R.}~\bibnamefont
  {G{\'o}mez-Bombarelli}},\ }\href
  {https://www.nature.com/articles/s41467-024-50407-9} {\bibfield  {journal}
  {\bibinfo  {journal} {Nature Communications}\ }\textbf {\bibinfo {volume}
  {15}},\ \bibinfo {pages} {6030} (\bibinfo {year} {2024})}\BibitemShut
  {NoStop}%
\bibitem [{\citenamefont {Zaverkin}\ \emph {et~al.}(2024)\citenamefont
  {Zaverkin}, \citenamefont {Holzm{\"u}ller}, \citenamefont {Christiansen},
  \citenamefont {Errica}, \citenamefont {Alesiani}, \citenamefont {Takamoto},
  \citenamefont {Niepert},\ and\ \citenamefont
  {K{\"a}stner}}]{zaverkin2024uncertainty}%
  \BibitemOpen
  \bibfield  {author} {\bibinfo {author} {\bibfnamefont {V.}~\bibnamefont
  {Zaverkin}}, \bibinfo {author} {\bibfnamefont {D.}~\bibnamefont
  {Holzm{\"u}ller}}, \bibinfo {author} {\bibfnamefont {H.}~\bibnamefont
  {Christiansen}}, \bibinfo {author} {\bibfnamefont {F.}~\bibnamefont
  {Errica}}, \bibinfo {author} {\bibfnamefont {F.}~\bibnamefont {Alesiani}},
  \bibinfo {author} {\bibfnamefont {M.}~\bibnamefont {Takamoto}}, \bibinfo
  {author} {\bibfnamefont {M.}~\bibnamefont {Niepert}}, \ and\ \bibinfo
  {author} {\bibfnamefont {J.}~\bibnamefont {K{\"a}stner}},\ }\href
  {https://www.nature.com/articles/s41524-024-01254-1} {\bibfield  {journal}
  {\bibinfo  {journal} {npj Computational Materials}\ }\textbf {\bibinfo
  {volume} {10}},\ \bibinfo {pages} {83} (\bibinfo {year} {2024})}\BibitemShut
  {NoStop}%
\bibitem [{\citenamefont {Palmer}\ \emph {et~al.}(2022)\citenamefont {Palmer},
  \citenamefont {Du}, \citenamefont {Politowicz}, \citenamefont {Emory},
  \citenamefont {Yang}, \citenamefont {Gautam}, \citenamefont {Gupta},
  \citenamefont {Li}, \citenamefont {Jacobs},\ and\ \citenamefont
  {Morgan}}]{palmer_calibration_2022}%
  \BibitemOpen
  \bibfield  {author} {\bibinfo {author} {\bibfnamefont {G.}~\bibnamefont
  {Palmer}}, \bibinfo {author} {\bibfnamefont {S.}~\bibnamefont {Du}}, \bibinfo
  {author} {\bibfnamefont {A.}~\bibnamefont {Politowicz}}, \bibinfo {author}
  {\bibfnamefont {J.~P.}\ \bibnamefont {Emory}}, \bibinfo {author}
  {\bibfnamefont {X.}~\bibnamefont {Yang}}, \bibinfo {author} {\bibfnamefont
  {A.}~\bibnamefont {Gautam}}, \bibinfo {author} {\bibfnamefont
  {G.}~\bibnamefont {Gupta}}, \bibinfo {author} {\bibfnamefont
  {Z.}~\bibnamefont {Li}}, \bibinfo {author} {\bibfnamefont {R.}~\bibnamefont
  {Jacobs}}, \ and\ \bibinfo {author} {\bibfnamefont {D.}~\bibnamefont
  {Morgan}},\ }\href {\doibase 10.1038/s41524-022-00794-8} {\bibfield
  {journal} {\bibinfo  {journal} {npj Computational Materials}\ }\textbf
  {\bibinfo {volume} {8}},\ \bibinfo {pages} {115} (\bibinfo {year}
  {2022})}\BibitemShut {NoStop}%
\bibitem [{\citenamefont {Musil}\ \emph {et~al.}(2019)\citenamefont {Musil},
  \citenamefont {Willatt}, \citenamefont {Langovoy},\ and\ \citenamefont
  {Ceriotti}}]{musil_fast_2019}%
  \BibitemOpen
  \bibfield  {author} {\bibinfo {author} {\bibfnamefont {F.}~\bibnamefont
  {Musil}}, \bibinfo {author} {\bibfnamefont {M.~J.}\ \bibnamefont {Willatt}},
  \bibinfo {author} {\bibfnamefont {M.~A.}\ \bibnamefont {Langovoy}}, \ and\
  \bibinfo {author} {\bibfnamefont {M.}~\bibnamefont {Ceriotti}},\ }\href
  {\doibase 10.1021/acs.jctc.8b00959} {\bibfield  {journal} {\bibinfo
  {journal} {Journal of Chemical Theory and Computation}\ }\textbf {\bibinfo
  {volume} {15}},\ \bibinfo {pages} {906} (\bibinfo {year} {2019})}\BibitemShut
  {NoStop}%
\bibitem [{\citenamefont {Imbalzano}\ \emph {et~al.}(2021)\citenamefont
  {Imbalzano}, \citenamefont {Zhuang}, \citenamefont {Kapil}, \citenamefont
  {Rossi}, \citenamefont {Engel}, \citenamefont {Grasselli},\ and\
  \citenamefont {Ceriotti}}]{imbalzano_uncertainty_2021}%
  \BibitemOpen
  \bibfield  {author} {\bibinfo {author} {\bibfnamefont {G.}~\bibnamefont
  {Imbalzano}}, \bibinfo {author} {\bibfnamefont {Y.}~\bibnamefont {Zhuang}},
  \bibinfo {author} {\bibfnamefont {V.}~\bibnamefont {Kapil}}, \bibinfo
  {author} {\bibfnamefont {K.}~\bibnamefont {Rossi}}, \bibinfo {author}
  {\bibfnamefont {E.~A.}\ \bibnamefont {Engel}}, \bibinfo {author}
  {\bibfnamefont {F.}~\bibnamefont {Grasselli}}, \ and\ \bibinfo {author}
  {\bibfnamefont {M.}~\bibnamefont {Ceriotti}},\ }\href
  {https://doi.org/10.1063/5.0036522} {\bibfield  {journal} {\bibinfo
  {journal} {The Journal of Chemical Physics}\ }\textbf {\bibinfo {volume}
  {154}},\ \bibinfo {pages} {074102} (\bibinfo {year} {2021})}\BibitemShut
  {NoStop}%
\bibitem [{\citenamefont {Bore}\ and\ \citenamefont
  {Paesani}(2023)}]{bore_realistic_2023}%
  \BibitemOpen
  \bibfield  {author} {\bibinfo {author} {\bibfnamefont {S.~L.}\ \bibnamefont
  {Bore}}\ and\ \bibinfo {author} {\bibfnamefont {F.}~\bibnamefont {Paesani}},\
  }\href {\doibase 10.1038/s41467-023-38855-1} {\bibfield  {journal} {\bibinfo
  {journal} {Nature Communications}\ }\textbf {\bibinfo {volume} {14}},\
  \bibinfo {pages} {3349} (\bibinfo {year} {2023})}\BibitemShut {NoStop}%
\bibitem [{\citenamefont {Dasgupta}\ \emph {et~al.}(2021)\citenamefont
  {Dasgupta}, \citenamefont {Lambros}, \citenamefont {Perdew},\ and\
  \citenamefont {Paesani}}]{dasgupta_elevating_2021}%
  \BibitemOpen
  \bibfield  {author} {\bibinfo {author} {\bibfnamefont {S.}~\bibnamefont
  {Dasgupta}}, \bibinfo {author} {\bibfnamefont {E.}~\bibnamefont {Lambros}},
  \bibinfo {author} {\bibfnamefont {J.~P.}\ \bibnamefont {Perdew}}, \ and\
  \bibinfo {author} {\bibfnamefont {F.}~\bibnamefont {Paesani}},\ }\href
  {\doibase 10.1038/s41467-021-26618-9} {\bibfield  {journal} {\bibinfo
  {journal} {Nature Communications}\ }\textbf {\bibinfo {volume} {12}},\
  \bibinfo {pages} {6359} (\bibinfo {year} {2021})}\BibitemShut {NoStop}%
\bibitem [{\citenamefont {Dasgupta}\ \emph {et~al.}(2024)\citenamefont
  {Dasgupta}, \citenamefont {Cassone},\ and\ \citenamefont
  {Paesani}}]{dasgupta_nuclear_2024}%
  \BibitemOpen
  \bibfield  {author} {\bibinfo {author} {\bibfnamefont {S.}~\bibnamefont
  {Dasgupta}}, \bibinfo {author} {\bibfnamefont {G.}~\bibnamefont {Cassone}}, \
  and\ \bibinfo {author} {\bibfnamefont {F.}~\bibnamefont {Paesani}},\ }\href
  {\doibase 10.26434/chemrxiv-2024-zkz7v} {\enquote {\bibinfo {title} {Nuclear
  quantum effects and the {Grotthuss} mechanism dictate the {pH} of liquid
  water},}\ } (\bibinfo {year} {2024})\BibitemShut {NoStop}%
\bibitem [{\citenamefont {Zhai}\ \emph {et~al.}(2024)\citenamefont {Zhai},
  \citenamefont {Rashmi}, \citenamefont {Palos},\ and\ \citenamefont
  {Paesani}}]{zhai2024many}%
  \BibitemOpen
  \bibfield  {author} {\bibinfo {author} {\bibfnamefont {Y.}~\bibnamefont
  {Zhai}}, \bibinfo {author} {\bibfnamefont {R.}~\bibnamefont {Rashmi}},
  \bibinfo {author} {\bibfnamefont {E.}~\bibnamefont {Palos}}, \ and\ \bibinfo
  {author} {\bibfnamefont {F.}~\bibnamefont {Paesani}},\ }\href
  {https://doi.org/10.1063/5.0203682} {\bibfield  {journal} {\bibinfo
  {journal} {The Journal of Chemical Physics}\ }\textbf {\bibinfo {volume}
  {160}} (\bibinfo {year} {2024})}\BibitemShut {NoStop}%
\bibitem [{\citenamefont {Maxson}\ and\ \citenamefont
  {Szilv{\'a}si}(2024)}]{maxson2024transferable}%
  \BibitemOpen
  \bibfield  {author} {\bibinfo {author} {\bibfnamefont {T.}~\bibnamefont
  {Maxson}}\ and\ \bibinfo {author} {\bibfnamefont {T.}~\bibnamefont
  {Szilv{\'a}si}},\ }\href {https://doi.org/10.1021/acs.jpclett.4c00605}
  {\bibfield  {journal} {\bibinfo  {journal} {The Journal of Physical Chemistry
  Letters}\ }\textbf {\bibinfo {volume} {15}},\ \bibinfo {pages} {3740}
  (\bibinfo {year} {2024})}\BibitemShut {NoStop}%
\bibitem [{\citenamefont {Cavka}\ \emph {et~al.}(2008)\citenamefont {Cavka},
  \citenamefont {Jakobsen}, \citenamefont {Olsbye}, \citenamefont {Guillou},
  \citenamefont {Lamberti}, \citenamefont {Bordiga},\ and\ \citenamefont
  {Lillerud}}]{cavka2008new}%
  \BibitemOpen
  \bibfield  {author} {\bibinfo {author} {\bibfnamefont {J.~H.}\ \bibnamefont
  {Cavka}}, \bibinfo {author} {\bibfnamefont {S.}~\bibnamefont {Jakobsen}},
  \bibinfo {author} {\bibfnamefont {U.}~\bibnamefont {Olsbye}}, \bibinfo
  {author} {\bibfnamefont {N.}~\bibnamefont {Guillou}}, \bibinfo {author}
  {\bibfnamefont {C.}~\bibnamefont {Lamberti}}, \bibinfo {author}
  {\bibfnamefont {S.}~\bibnamefont {Bordiga}}, \ and\ \bibinfo {author}
  {\bibfnamefont {K.~P.}\ \bibnamefont {Lillerud}},\ }\href
  {https://doi.org/10.1021/ja8057953} {\bibfield  {journal} {\bibinfo
  {journal} {Journal of the American Chemical Society}\ }\textbf {\bibinfo
  {volume} {130}},\ \bibinfo {pages} {13850} (\bibinfo {year}
  {2008})}\BibitemShut {NoStop}%
\bibitem [{\citenamefont {Virtanen}\ \emph {et~al.}(2020)\citenamefont
  {Virtanen} \emph {et~al.}}]{virtanen_scipy_2020}%
  \BibitemOpen
  \bibfield  {author} {\bibinfo {author} {\bibfnamefont {P.}~\bibnamefont
  {Virtanen}} \emph {et~al.},\ }\href {\doibase 10.1038/s41592-019-0686-2}
  {\bibfield  {journal} {\bibinfo  {journal} {Nature Methods}\ }\textbf
  {\bibinfo {volume} {17}},\ \bibinfo {pages} {261} (\bibinfo {year}
  {2020})}\BibitemShut {NoStop}%
\bibitem [{\citenamefont {Zeng}\ \emph {et~al.}(2023)\citenamefont {Zeng} \emph
  {et~al.}}]{zeng_deepmd-kit_2023}%
  \BibitemOpen
  \bibfield  {author} {\bibinfo {author} {\bibfnamefont {J.}~\bibnamefont
  {Zeng}} \emph {et~al.},\ }\href {\doibase 10.1063/5.0155600} {\bibfield
  {journal} {\bibinfo  {journal} {The Journal of Chemical Physics}\ }\textbf
  {\bibinfo {volume} {159}},\ \bibinfo {pages} {054801} (\bibinfo {year}
  {2023})}\BibitemShut {NoStop}%
\bibitem [{\citenamefont {Sciortino}\ \emph {et~al.}(2025)\citenamefont
  {Sciortino}, \citenamefont {Zhai}, \citenamefont {Bore},\ and\ \citenamefont
  {Paesani}}]{sciortino2025constraints}%
  \BibitemOpen
  \bibfield  {author} {\bibinfo {author} {\bibfnamefont {F.}~\bibnamefont
  {Sciortino}}, \bibinfo {author} {\bibfnamefont {Y.}~\bibnamefont {Zhai}},
  \bibinfo {author} {\bibfnamefont {S.}~\bibnamefont {Bore}}, \ and\ \bibinfo
  {author} {\bibfnamefont {F.}~\bibnamefont {Paesani}},\ }\href
  {https://www.nature.com/articles/s41567-024-02761-0} {\bibfield  {journal}
  {\bibinfo  {journal} {Nature Physics}\ ,\ \bibinfo {pages} {1}} (\bibinfo
  {year} {2025})}\BibitemShut {NoStop}%
\bibitem [{\citenamefont {Babin}\ \emph {et~al.}(2013)\citenamefont {Babin},
  \citenamefont {Leforestier},\ and\ \citenamefont
  {Paesani}}]{babin2013development}%
  \BibitemOpen
  \bibfield  {author} {\bibinfo {author} {\bibfnamefont {V.}~\bibnamefont
  {Babin}}, \bibinfo {author} {\bibfnamefont {C.}~\bibnamefont {Leforestier}},
  \ and\ \bibinfo {author} {\bibfnamefont {F.}~\bibnamefont {Paesani}},\ }\href
  {https://doi.org/10.1021/ct400863t} {\bibfield  {journal} {\bibinfo
  {journal} {Journal of Chemical Theory and Computation}\ }\textbf {\bibinfo
  {volume} {9}},\ \bibinfo {pages} {5395} (\bibinfo {year} {2013})}\BibitemShut
  {NoStop}%
\bibitem [{\citenamefont {Babin}\ \emph {et~al.}(2014)\citenamefont {Babin},
  \citenamefont {Medders},\ and\ \citenamefont
  {Paesani}}]{babin2014development}%
  \BibitemOpen
  \bibfield  {author} {\bibinfo {author} {\bibfnamefont {V.}~\bibnamefont
  {Babin}}, \bibinfo {author} {\bibfnamefont {G.~R.}\ \bibnamefont {Medders}},
  \ and\ \bibinfo {author} {\bibfnamefont {F.}~\bibnamefont {Paesani}},\ }\href
  {https://doi.org/10.1021/ct500079y} {\bibfield  {journal} {\bibinfo
  {journal} {Journal of Chemical Theory and Computation}\ }\textbf {\bibinfo
  {volume} {10}},\ \bibinfo {pages} {1599} (\bibinfo {year}
  {2014})}\BibitemShut {NoStop}%
\bibitem [{\citenamefont {Belleflamme}\ and\ \citenamefont
  {Hutter}(2023)}]{belleflamme_radicals_2023}%
  \BibitemOpen
  \bibfield  {author} {\bibinfo {author} {\bibfnamefont {F.}~\bibnamefont
  {Belleflamme}}\ and\ \bibinfo {author} {\bibfnamefont {J.}~\bibnamefont
  {Hutter}},\ }\href {\doibase 10.1039/D3CP02517A} {\bibfield  {journal}
  {\bibinfo  {journal} {Physical Chemistry Chemical Physics}\ }\textbf
  {\bibinfo {volume} {25}},\ \bibinfo {pages} {20817} (\bibinfo {year}
  {2023})}\BibitemShut {NoStop}%
\bibitem [{hsp(2025)}]{hsp}%
  \BibitemOpen
  \href {https://gitlab.com/hylleraasplatform} {\bibfield  {journal} {\bibinfo
  {journal} {Hylleraas Software Platform}\ }\textbf {\bibinfo {volume}
  {hyal}},\ \bibinfo {pages} {https://gitlab.com/hylleraasplatform} (\bibinfo
  {year} {2025})}\BibitemShut {NoStop}%
\bibitem [{\citenamefont {Thompson}\ \emph {et~al.}(2022)\citenamefont
  {Thompson}, \citenamefont {Aktulga}, \citenamefont {Berger}, \citenamefont
  {Bolintineanu}, \citenamefont {Brown}, \citenamefont {Crozier}, \citenamefont
  {In~'T~Veld}, \citenamefont {Kohlmeyer}, \citenamefont {Moore}, \citenamefont
  {Nguyen}, \citenamefont {Shan}, \citenamefont {Stevens}, \citenamefont
  {Tranchida}, \citenamefont {Trott},\ and\ \citenamefont
  {Plimpton}}]{thompson_lammps_2022}%
  \BibitemOpen
  \bibfield  {author} {\bibinfo {author} {\bibfnamefont {A.~P.}\ \bibnamefont
  {Thompson}}, \bibinfo {author} {\bibfnamefont {H.~M.}\ \bibnamefont
  {Aktulga}}, \bibinfo {author} {\bibfnamefont {R.}~\bibnamefont {Berger}},
  \bibinfo {author} {\bibfnamefont {D.~S.}\ \bibnamefont {Bolintineanu}},
  \bibinfo {author} {\bibfnamefont {W.~M.}\ \bibnamefont {Brown}}, \bibinfo
  {author} {\bibfnamefont {P.~S.}\ \bibnamefont {Crozier}}, \bibinfo {author}
  {\bibfnamefont {P.~J.}\ \bibnamefont {In~'T~Veld}}, \bibinfo {author}
  {\bibfnamefont {A.}~\bibnamefont {Kohlmeyer}}, \bibinfo {author}
  {\bibfnamefont {S.~G.}\ \bibnamefont {Moore}}, \bibinfo {author}
  {\bibfnamefont {T.~D.}\ \bibnamefont {Nguyen}}, \bibinfo {author}
  {\bibfnamefont {R.}~\bibnamefont {Shan}}, \bibinfo {author} {\bibfnamefont
  {M.~J.}\ \bibnamefont {Stevens}}, \bibinfo {author} {\bibfnamefont
  {J.}~\bibnamefont {Tranchida}}, \bibinfo {author} {\bibfnamefont
  {C.}~\bibnamefont {Trott}}, \ and\ \bibinfo {author} {\bibfnamefont {S.~J.}\
  \bibnamefont {Plimpton}},\ }\href {\doibase 10.1016/j.cpc.2021.108171}
  {\bibfield  {journal} {\bibinfo  {journal} {Computer Physics Communications}\
  }\textbf {\bibinfo {volume} {271}},\ \bibinfo {pages} {108171} (\bibinfo
  {year} {2022})}\BibitemShut {NoStop}%
\end{thebibliography}%


\begin{thebibliography}{6}%
\makeatletter
\providecommand \@ifxundefined [1]{%
 \@ifx{#1\undefined}
}%
\providecommand \@ifnum [1]{%
 \ifnum #1\expandafter \@firstoftwo
 \else \expandafter \@secondoftwo
 \fi
}%
\providecommand \@ifx [1]{%
 \ifx #1\expandafter \@firstoftwo
 \else \expandafter \@secondoftwo
 \fi
}%
\providecommand \natexlab [1]{#1}%
\providecommand \enquote  [1]{``#1''}%
\providecommand \bibnamefont  [1]{#1}%
\providecommand \bibfnamefont [1]{#1}%
\providecommand \citenamefont [1]{#1}%
\providecommand \href@noop [0]{\@secondoftwo}%
\providecommand \href [0]{\begingroup \@sanitize@url \@href}%
\providecommand \@href[1]{\@@startlink{#1}\@@href}%
\providecommand \@@href[1]{\endgroup#1\@@endlink}%
\providecommand \@sanitize@url [0]{\catcode `\\12\catcode `\$12\catcode
  `\&12\catcode `\#12\catcode `\^12\catcode `\_12\catcode `\%12\relax}%
\providecommand \@@startlink[1]{}%
\providecommand \@@endlink[0]{}%
\providecommand \url  [0]{\begingroup\@sanitize@url \@url }%
\providecommand \@url [1]{\endgroup\@href {#1}{\urlprefix }}%
\providecommand \urlprefix  [0]{URL }%
\providecommand \Eprint [0]{\href }%
\providecommand \doibase [0]{http://dx.doi.org/}%
\providecommand \selectlanguage [0]{\@gobble}%
\providecommand \bibinfo  [0]{\@secondoftwo}%
\providecommand \bibfield  [0]{\@secondoftwo}%
\providecommand \translation [1]{[#1]}%
\providecommand \BibitemOpen [0]{}%
\providecommand \bibitemStop [0]{}%
\providecommand \bibitemNoStop [0]{.\EOS\space}%
\providecommand \EOS [0]{\spacefactor3000\relax}%
\providecommand \BibitemShut  [1]{\csname bibitem#1\endcsname}%
\let\auto@bib@innerbib\@empty
\bibitem [{\citenamefont {Bore}\ and\ \citenamefont
  {Paesani}(2023)}]{bore_realistic_2023}%
  \BibitemOpen
  \bibfield  {author} {\bibinfo {author} {\bibfnamefont {S.~L.}\ \bibnamefont
  {Bore}}\ and\ \bibinfo {author} {\bibfnamefont {F.}~\bibnamefont {Paesani}},\
  }\href {\doibase 10.1038/s41467-023-38855-1} {\bibfield  {journal} {\bibinfo
  {journal} {Nature Communications}\ }\textbf {\bibinfo {volume} {14}},\
  \bibinfo {pages} {3349} (\bibinfo {year} {2023})}\BibitemShut {NoStop}%
\bibitem [{\citenamefont {Mahoney}\ and\ \citenamefont
  {Jorgensen}(2000)}]{mahoney-fivesite-2000}%
  \BibitemOpen
  \bibfield  {author} {\bibinfo {author} {\bibfnamefont {M.~W.}\ \bibnamefont
  {Mahoney}}\ and\ \bibinfo {author} {\bibfnamefont {W.~L.}\ \bibnamefont
  {Jorgensen}},\ }\href {\doibase 10.1063/1.481505} {\bibfield  {journal}
  {\bibinfo  {journal} {The Journal of Chemical Physics}\ }\textbf {\bibinfo
  {volume} {112}},\ \bibinfo {pages} {8910–8922} (\bibinfo {year}
  {2000})}\BibitemShut {NoStop}%
\bibitem [{\citenamefont {{Materials Design Group}}(2025)}]{wmd-crystal-2025}%
  \BibitemOpen
  \bibfield  {author} {\bibinfo {author} {\bibnamefont {{Materials Design
  Group}}},\ }\href@noop {} {\enquote {\bibinfo {title} {Crystal structures},}\
  }\bibinfo {howpublished}
  {\url{https://github.com/WMD-group/Crystal_structures/}} (\bibinfo {year}
  {2025})\BibitemShut {NoStop}%
\bibitem [{\citenamefont {Martínez}\ \emph {et~al.}(2009)\citenamefont
  {Martínez}, \citenamefont {Andrade}, \citenamefont {Birgin},\ and\
  \citenamefont {Martínez}}]{martinez_packmol_2009}%
  \BibitemOpen
  \bibfield  {author} {\bibinfo {author} {\bibfnamefont {L.}~\bibnamefont
  {Martínez}}, \bibinfo {author} {\bibfnamefont {R.}~\bibnamefont {Andrade}},
  \bibinfo {author} {\bibfnamefont {E.~G.}\ \bibnamefont {Birgin}}, \ and\
  \bibinfo {author} {\bibfnamefont {J.~M.}\ \bibnamefont {Martínez}},\ }\href
  {\doibase 10.1002/jcc.21224} {\bibfield  {journal} {\bibinfo  {journal}
  {Journal of Computational Chemistry}\ }\textbf {\bibinfo {volume} {30}},\
  \bibinfo {pages} {2157–2164} (\bibinfo {year} {2009})}\BibitemShut
  {NoStop}%
\bibitem [{\citenamefont {Perdew}\ \emph {et~al.}(1996)\citenamefont {Perdew},
  \citenamefont {Burke},\ and\ \citenamefont
  {Ernzerhof}}]{perdew-generalized-1996}%
  \BibitemOpen
  \bibfield  {author} {\bibinfo {author} {\bibfnamefont {J.~P.}\ \bibnamefont
  {Perdew}}, \bibinfo {author} {\bibfnamefont {K.}~\bibnamefont {Burke}}, \
  and\ \bibinfo {author} {\bibfnamefont {M.}~\bibnamefont {Ernzerhof}},\ }\href
  {\doibase 10.1103/physrevlett.77.3865} {\bibfield  {journal} {\bibinfo
  {journal} {Physical Review Letters}\ }\textbf {\bibinfo {volume} {77}},\
  \bibinfo {pages} {3865–3868} (\bibinfo {year} {1996})}\BibitemShut
  {NoStop}%
\bibitem [{\citenamefont {Grimme}\ \emph {et~al.}(2010)\citenamefont {Grimme},
  \citenamefont {Antony}, \citenamefont {Ehrlich},\ and\ \citenamefont
  {Krieg}}]{grimme-consistent-2010}%
  \BibitemOpen
  \bibfield  {author} {\bibinfo {author} {\bibfnamefont {S.}~\bibnamefont
  {Grimme}}, \bibinfo {author} {\bibfnamefont {J.}~\bibnamefont {Antony}},
  \bibinfo {author} {\bibfnamefont {S.}~\bibnamefont {Ehrlich}}, \ and\
  \bibinfo {author} {\bibfnamefont {H.}~\bibnamefont {Krieg}},\ }\href
  {\doibase 10.1063/1.3382344} {\bibfield  {journal} {\bibinfo  {journal} {The
  Journal of Chemical Physics}\ }\textbf {\bibinfo {volume} {132}} (\bibinfo
  {year} {2010}),\ 10.1063/1.3382344}\BibitemShut {NoStop}%
\end{thebibliography}%

\end{document}


\title{Supplementary Information for:\\
Learning atomic forces from uncertainty-calibrated adversarial attacks}
\author{Henrique Musseli Cezar}
\affiliation{Hylleraas Centre for Quantum Molecular Sciences and Department of Chemistry, University of Oslo, PO Box 1033 Blindern, 0315 Oslo, Norway}

\author{Tilmann Bodenstein}
\affiliation{Hylleraas Centre for Quantum Molecular Sciences and Department of Chemistry, University of Oslo, PO Box 1033 Blindern, 0315 Oslo, Norway}

\author{Henrik Andersen Sveinsson}
\affiliation{The Njord Centre, Department of Physics, University of Oslo, PO Box 1048 Blindern, 0316 Oslo, Norway}

\author{Morten Ledum}
\affiliation{Hylleraas Centre for Quantum Molecular Sciences and Department of Chemistry, University of Oslo, PO Box 1033 Blindern, 0315 Oslo, Norway}

\author{Simen Reine}
\affiliation{Hylleraas Centre for Quantum Molecular Sciences and Department of Chemistry, University of Oslo, PO Box 1033 Blindern, 0315 Oslo, Norway}

\author{Sigbjørn Løland Bore}
\email{s.l.bore@kjemi.uio.no}
\affiliation{Hylleraas Centre for Quantum Molecular Sciences and Department of Chemistry, University of Oslo, PO Box 1033 Blindern, 0315 Oslo, Norway}
\maketitle

{\small
\newpage
\tableofcontents
\newpage
}

\section{Additional analysis for CAGO to target error}
\begin{figure}[!htp]
    \centering
    \includegraphics{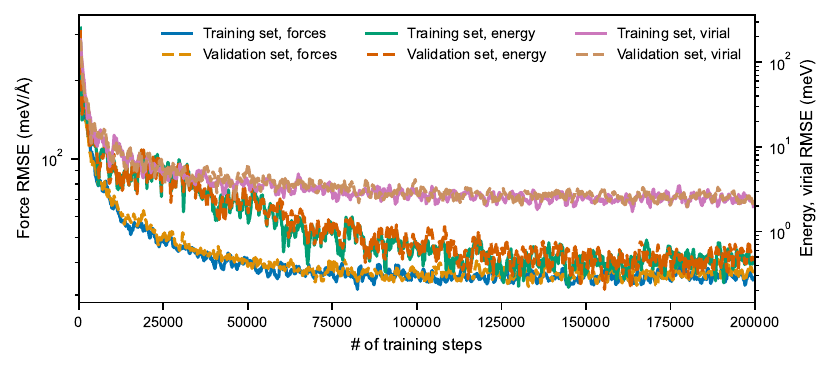}
    \caption{Representative learning curve for forces, energies, and virials for one out of 20 MLIP models for DNN@MB-pol, with a running average over ten training steps.}
    \label{sm:fig:lcurve}
\end{figure}

Training, validation, and test sets were created as specified in Table~\ref{sm:tab:train-val-test}. Twenty models were trained for 200000 steps, with an example learning curve for a single model shown in Figure~\ref{sm:fig:lcurve}. Then, picking out 40 structures from the test set, we apply the adversarial geometry optimization on those structures to benchmark different settings, such as uncertainty calibration, committee size, etc. In the final geometries, we compute the reference DNN@MB-pol forces, which we use to compute the errors for the committee.

\begin{table}[!htb]
  \centering
  \caption{The training, validation, and test set were created from DNN@MB-pol MD simulations of liquid water with 32 water molecules. The structures were gathered from 0.5~ns trajectories at various temperatures, disregarding the first 0.1~ns, for two different seeds each. For each of the 20 models, we created individual bootstrap training and validation sets from simulations with seed 1. The test set was created from simulations of seed 2, thereby being distinct from training and validation sets.}\label{sm:tab:train-val-test}
\begin{tabular}{ccc}
\hline\hline
Type&\# of structures& source\\
\hline
     Training set&500& 1~Bar, [300~K, 350~K, 400~K, 450~K, 500~K], seed 1\\
     Validation set&500&1~Bar, [300~K, 350~K, 400~K, 450~K, 500~K], seed 1\\
     Test set & 1000 &1~Bar, [300~K, 350~K, 400~K, 450~K, 500~K], seed 2\\ \hline\hline
\end{tabular}
\end{table}  

\begin{figure}[!htb]
    \centering
    \includegraphics[width=1\textwidth]{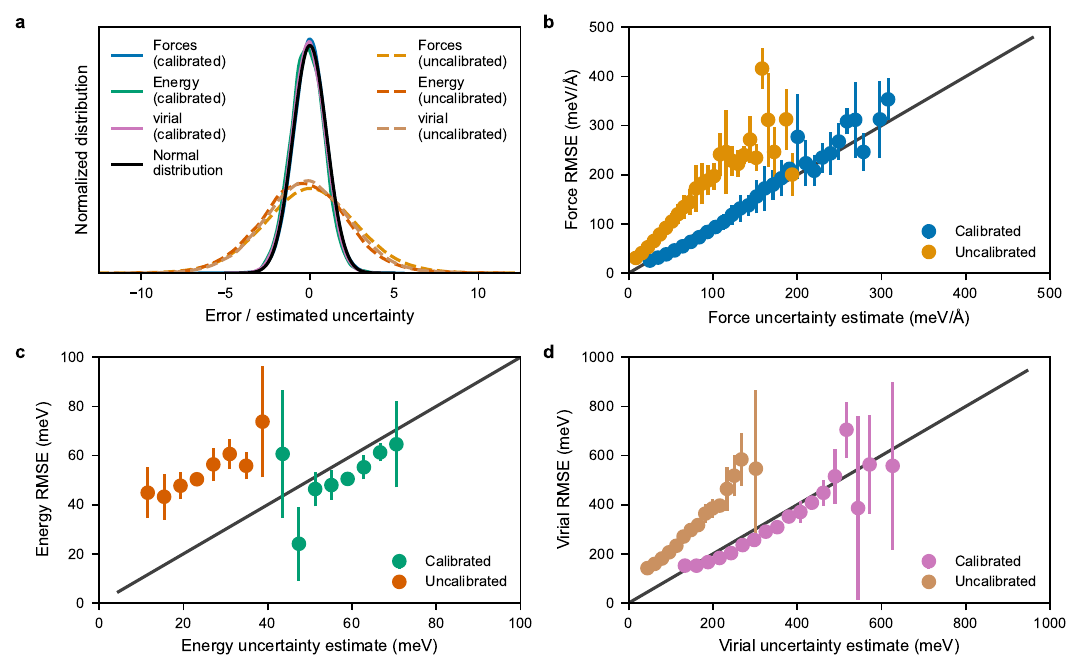}
    \caption{Statistical properties of calibrated uncertainty. (a) The distribution of errors is divided by estimated uncertainty for force, energy, and virial prediction, with and without error calibration. (b,c,d)  Average RMSE binned by uncertainty estimate for calibrated and uncalibrated uncertainty for forces, energy and virials, respectively. Error bars correspond to 95~\% confidence intervals. }
    \label{fig:sm:calibration}
\end{figure}

Our negative log-likelihood optimization determines the uncertainty calibrations reported in Table~\ref{sm:tab:calibration-parameters}. Figure~\ref{fig:sm:calibration}a reports the distribution residuals, error divided by estimated uncertainty for forces, energies, and virials. For well-calibrated uncertainty estimates, residuals are distributed according to the normal distribution with mean zero and standard deviation equal to 1. This is the case for calibrated uncertainty for energy, forces, and virials, while the uncalibrated uncertainty underestimates the error, leading to broader distributions. The same picture is reflected in our force RMSE binned over uncertainty estimate in Figure~\ref{fig:sm:calibration}b, c, and d, where calibrated uncertainty has a near-perfect correlation with force RMSE, virial and energy RMSE. In contrast, uncalibrated uncertainty underestimates the true error.  
\begin{table}[!htb]
    \centering
        \caption{Calibration parameters for power-law correction.}\label{sm:tab:calibration-parameters}
\begin{tabular}{lccc}
\hline\hline
Property & Unit & a & b \\
\hline
forces & eV/Å & 0.609 & 0.506 \\
energy & eV & 0.8799& 0.7379 \\
virial & eV & 1.338& 0.7256 \\
\hline\hline
\end{tabular}
\end{table}

\begin{figure}[!htb]
    \centering
    \includegraphics[width=1\textwidth]{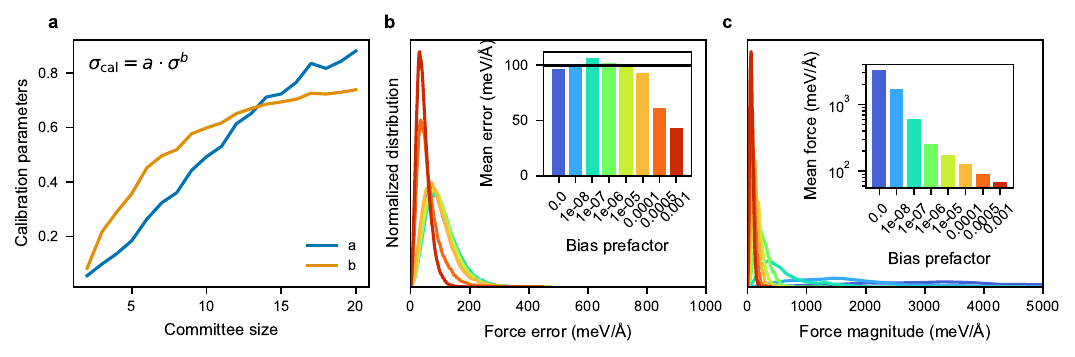}
    \caption{Statistical properties of calibrated uncertainty. \textbf{a.} Uncertainty calibration constants as a function of committee size. \textbf{b.} Distribution of prediction error for different force bias prefactors and corresponding mean force error. \textbf{c.} Distribution of force magnitude for different force-bias prefactors, as well as from reference molecular dynamics simulations and corresponding means in inset. }
    \label{fig:sm:benchmark}
\end{figure}

As discussed in the main text, the size of the committees affects the calibration, as can be seen from the change of the calibration parameters with respect to committee size in Figure~\ref{fig:sm:benchmark}a. However, despite the calibration parameters being different for the different committee sizes, as shown in the main text, starting from 10 committee members, the calibration is sufficient to provide reliable results. In Figure~\ref{fig:sm:benchmark}b and c, we show an example of biasing towards zero average force magnitude. If chosen moderately, one can avoid highly steric structures, with error statistics largely unaffected. This is potentially useful when subsequently doing reference calculations, which can either fail or struggle to converge for structures dominated by steric repulsions.

\section{Structures and training sets}
\begin{figure}[!htb]
    \centering
    \includegraphics[width=1\textwidth]{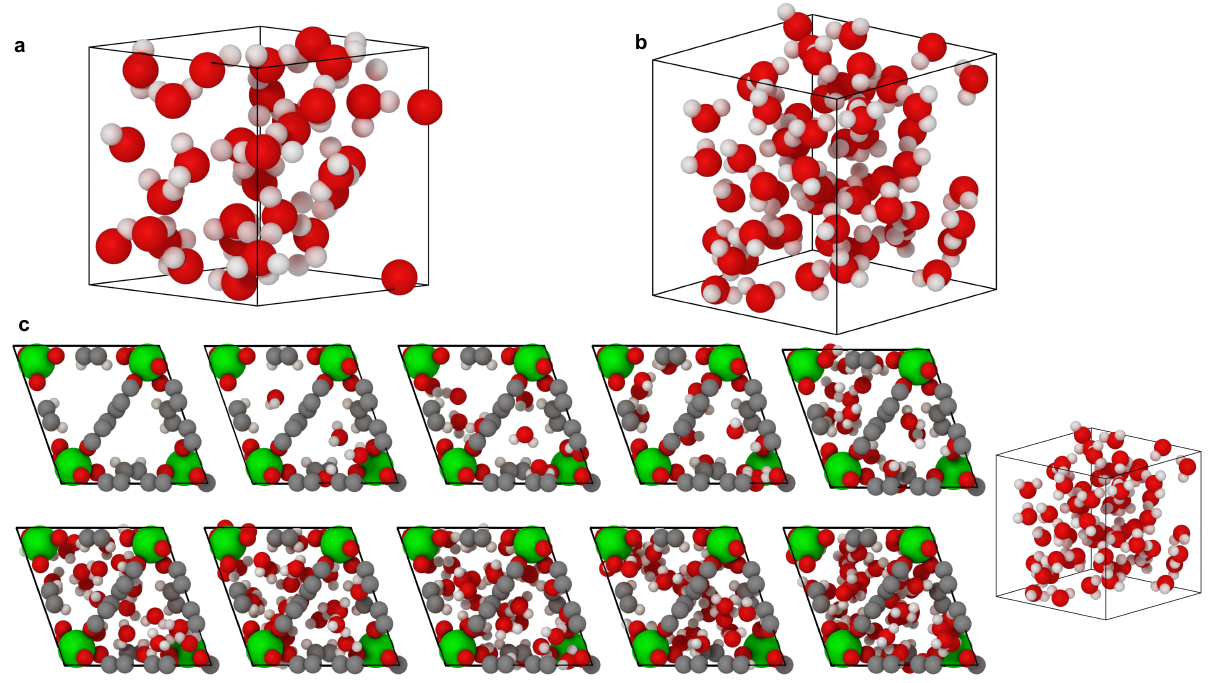}
    \caption{Molecular structures. {\bf a.} System I: box of 32 water molecules. {\bf b.} System II: Box of 64 water molecules. {\bf c.} System III: UiO-66 with water weight percentage from 0 to 45, and a box of 64 water molecules. The color coding for the atoms is red, white, gray, and green, respectively, for oxygen, hydrogen, carbon, and zirconium.}
    \label{sm-fig:systems}
\end{figure}

The main systems used to get the results presented in the manuscript are presented in Figure~\ref{sm-fig:systems}.
System I uses structures extracted from molecular dynamics simulations using DNN@MB-pol.\cite{bore_realistic_2023}
System II is created based on a snapshot of a TIP5P\cite{mahoney-fivesite-2000} simulation at 1~bar and 300 ~K.
System III is built starting from a crystal structure of UiO-66 optimized with PBEsol,\cite{wmd-crystal-2025} and by adding water molecules using Packmol.\cite{martinez_packmol_2009}
For the structures in systems II and III, we have geometry optimized the structures using the PBE functional\cite{perdew-generalized-1996} with Grimme dispersion corrections (DFT-D3)\cite{grimme-consistent-2010} before using them in the active learning loop. All the initial structures, as well as the CAGO optimized structures during the active learning loops, are provided in the data repository associated with this manuscript.

\section{Additional analysis for Learning liquid water from a single structure}
\begin{figure*}[!htb]
    \centering
    \includegraphics[width=0.99\linewidth]{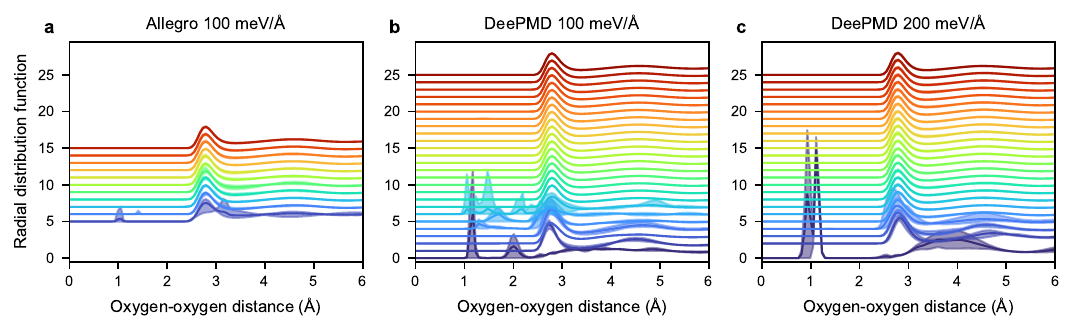}
    \caption{Oxygen-oxygen radial distribution functions for the pure water models. The radial distribution functions are shifted based on the iteration of the active learning loop, meaning that the radial distribution function that has its zero at 5, for example, belongs to the fifth iteration. The radial distribution functions are computed as the average between all the committee models, and the minimum and maximum value among the models for each distance is represented as the shaded region.}
    \label{sm:fig:rdfs-water}
\end{figure*}

In Figure~4 of the main text, we have computed the first coordination number based on the O-O radial distribution function. Here, in Figure~\ref{sm:fig:rdfs-water}, the radial distribution functions considering the average between all frames and models at each iteration for the DeePMD and Allegro models are shown.
The shaded regions in each color represent the minimum and maximum at each bin distance $r$ among the 12 different models of the committee.
In the first iterations, the models do not capture the liquid structure of water, especially the DeePMD models.
Many models also show peaks at distances shorter than the typical first peak of solvation around 2.5~\AA. As more configurations are added to the training set, the radial distribution functions converge to a liquid radial distribution function,  exhibiting well-converged radial distribution functions in the last few iterations. However, these small differences are captured by the first coordination number, as shown in Figure~4 of the main text.

\clearpage

\begin{table}[!htb]
\caption{Comparison of average force errors and water properties for the final models of the different active learning approaches reported in Figure~4.}
\label{tab:water_models}
{\small
\begin{tabular}{lccccc}
\toprule
Active learning approach & Mean force error & Density & Diffusion & First coordination \#\\
& meV/\AA  &g/cm$^3$ & $10^{-9}$ m$^2$/s &
\\
\midrule
Allegro CAGO
100 meV/Å  & 36.9 & 0.950 ± 0.009  & 1.586 ± 0.093   & 4.559 ± 0.092 \\
DeePMD CAGO 100 meV/Å & 63.0 & 0.943 ± 0.002  & 1.757 ± 0.082  & 4.475 ± 0.021 \\
DeePMD CAGO 200 meV/Å  & 70.3  & 0.944 ± 0.007  & 1.490 ± 0.112   & 4.450 ± 0.068 \\
DeePMD Max. uncertainty & 84.6  & 0.927 ± 0.004  & 1.185 ± 0.119   & 4.373 ± 0.035 \\
DeePMD Random 500 K & 76.7  & 0.937 ± 0.007  & 1.919 ± 0.149   & 4.401 ± 0.089 \\
\bottomrule
\end{tabular}
}
\end{table}
\bibliographystyle{apsrev4-1}
\bibliography{mybib}